\documentclass[11pt]{article}
\usepackage{amsmath}
\usepackage{amsmath}
\usepackage{amssymb}
\usepackage{physics}
\usepackage{cite}
\usepackage{blindtext} 
\usepackage{bookman}
\usepackage[T1]{fontenc}
\usepackage{microtype}
\usepackage[english]{babel} 
\usepackage[hmarginratio=1:1,top=32mm,columnsep=20pt]{geometry} 
\usepackage[hang, small,labelfont=bf,up,textfont=rm,up]{caption} 
\usepackage{booktabs} 
\usepackage{lettrine} 
\usepackage{enumitem} 
\setlist[itemize]{noitemsep} 
\usepackage{abstract} 
\usepackage{titlesec} 
\titlespacing*{\section}
{0pt}{1\baselineskip}{1\baselineskip}
\titlespacing*{\subsection}{0pt}{1\baselineskip}{1\baselineskip}
\titleformat{\section}[block]{\large\scshape\centering}{\thesection.}{1em}{} 
\titleformat{\subsection}[block]{\large}{\thesubsection.}{1em}{}
\usepackage{fancyhdr} 
\usepackage{titling} 
\usepackage{hyperref} 
\hypersetup{
    colorlinks=true,
    linkcolor=blue,
    filecolor=magenta,      
    urlcolor=blue,
    citecolor=blue,
    pdftitle={Overleaf Example},
    pdfpagemode=FullScreen,
    }

\usepackage[dvipsnames]{xcolor}
\usepackage[affil-it]{authblk}
\usepackage{amssymb,physics,slashed}
\usepackage{graphicx}
\title{QCD and electroweak phase transitions with hidden scale invariance: implications for primordial black holes, quark-lepton nuggets and gravitational waves. 
} 
\author{
\textsc{Joshua Cesca and Archil Kobakhidze} 
\vspace{0.2cm} \\
\normalsize \itshape
Sydney Consortium for Particle Physics and Cosmology, \\
\normalsize  \itshape
School of Physics, The University of Sydney, NSW 2006, Australia \\
\normalsize \itshape 
 \href{mailto:joshua.cesca@sydney.edu.au}{joshua.cesca@sydney.edu.au}, 
\href{mailto:archil.kobakhidze@sydney.edu.au}{archil.kobakhidze@sydney.edu.au}.

}
\date{} 
\renewcommand{\maketitlehookd}{

\begin{abstract}
\noindent 
We study the cosmological implications of the minimal non-linear realisation of scale invariance within the Standard Model (SM). This framework provides a technically natural explanation for the hierarchy between the Planck scale and the electroweak scale and introduces only a light, feebly coupled dilaton field beyond the SM particles. Although the model is almost indistinguishable from the minimal SM at low energies, its cosmological consequences differ dramatically. In particular, the electroweak Higgs field remains trapped in the symmetric phase until the Universe cools to very low temperatures, $T_c^{(\chi)}\sim 28$ MeV, where the first-order QCD chiral symmetry-breaking phase transition triggers the electroweak phase transition. This scenario offers intriguing possibilities for the production of primordial black holes, low-frequency gravitational waves, and multi-quark and lepton nuggets, which we explore in some detail using simplified approximations.
\end{abstract}
}

\begin{document}
\maketitle

\section{Introduction} 

Classical scale invariance has been proposed as a symmetry that could underlie the technically natural hierarchy of physical scales in particle physics \cite{Kobakhidze:2014afa, Meissner:2006zh}. A minimal implementation of hidden scale invariance within the Standard Model (SM) \cite{Kobakhidze:2017eml} predicts the existence of an additional light, feebly interacting particle: the dilaton.\footnote{The dilaton, a generic prediction of string theory, has also been argued to be an essential ingredient of any theory that aims to preserve the consistency of the low-energy ground state of general relativity \cite{Dvali:2024dlb}.} Owing to these properties, the resulting theory remains very close to the SM, making the two practically indistinguishable in collider experiments.

Nevertheless, the cosmological history of the universe gets dramatically modified \cite{Arunasalam:2017ajm}. The electroweak phase transition, by itself, cannot proceed efficiently. Instead, it requires assistance from the QCD chiral phase transition, which operates through the formation of quark condensates \cite{Witten:1980ez}. In this scenario, the electroweak Higgs field initially becomes trapped in its symmetric vacuum state. This trapping occurs because the scalar potential, being nearly scale-invariant, features a large potential barrier that prevents the transition to the broken-symmetry phase. As the universe expands and cools, quarks undergo hadronisation, during which they bind into hadrons and develop chiral-symmetry-breaking condensates. These condensates play a crucial role: through Higgs–Yukawa interactions, they destabilise the symmetric Higgs vacuum, thereby releasing the field from the barrier and driving the electroweak phase transition.

This mechanism constitutes a significant deviation from the conventional cosmological paradigm describing the early universe. Its implications are far-reaching, as it provides a natural pathway for the formation of primordial black holes (PBHs) and the production of low-frequency gravitational waves \cite{Arunasalam:2017ajm}. Moreover, the same theoretical framework offers a compelling basis for cold baryogenesis, thereby furnishing a potential explanation for the observed matter–antimatter asymmetry  \cite{Krauss:1999ng, Garcia-Bellido:1999xos}.

In this paper, we aim to construct a model of the QCD-triggered electroweak phase transition within a framework that incorporates hidden scale invariance \cite{Kobakhidze:2017eml}, and to compute the key features of this transition. The dynamics of such a transition are rather complicated, so we make a few simplified approximations based on the cosmic chronology and hierarchy of scales. As all six flavours of quarks (and other SM particles) remain massless prior to the electroweak symmetry breaking, QCD enters the strong coupling regime at a significantly smaller energy scale of the order of $\sim 90$ MeV. Hence, it is reasonable to assume that the hadronisation phase transition, accompanied by gluon condensates, proceeds first, once the universe cools down to temperatures $T_c^{(h)}\lesssim 90$ MeV.        

After the hadronisation phase, the dynamics are governed by an approximately classically scale-invariant scalar potential that couples the Higgs field ($h$), the composite sigma meson ($\sigma$), an extended sector of pions, and the dilaton. By solving for the dilaton field, we identify an unstable direction in the $h-\sigma$ field space. This instability triggers the chiral phase transition through the nucleation of vacuum bubbles in this unstable direction, followed by their expansion and eventual percolation.

We find that the transition completes at even lower temperatures, around $T_c^{(\chi)}\sim 28$ MeV, due to supercooling associated with the approximately scale-invariant form of the effective scalar potential. The chiral phase transition effectively removes the potential barrier for the Higgs field, allowing it to roll down to its vacuum expectation value, thereby completing the electroweak phase transition.

Our results, therefore, suggest that both the QCD and electroweak phase transitions occur only after the universe has cooled substantially. Within the approximations used, they also quantify and support the preliminary findings of Ref. \cite{Arunasalam:2017ajm}. Namely, the first-order nature of the phase transition assists the formation of primordial black holes. The matter can collapse into a black hole due to the overdensities produced during the phase transition, or due to the pre-existing (inflationary) overdensities. We find that the characteristics of the phase transition are such that the first scenario is highly unlikely to be realised. On the other hand, if overdensities were produced during the inflationary era, they could efficiently collapse into black holes of mass $m_{\rm PBH}\sim 1~M_{\odot}$ during the confinement phase transition at $T_c^{(h)}\sim 90$ MeV or $m_{\rm PBH}\sim 50~M_{\odot}$ during the chiral-electroweak phase transition occurring at $T_c^{(\chi)}\sim 28$ MeV. The abundance of such black holes is constrained to be $f_{\rm PBH}=\Omega_{\rm PBH}/\Omega_{\rm dm}\lesssim 10^{-2}-10^{-3}$ of the dark matter abundance. The production of this amount of primordial black holes requires a special arrangement of the dynamics during inflation. Namely, the amplitude of curvature perturbations at relevant small scales is required to be enhanced by $\sim 3-4$ orders of magnitude relative to the amplitude of perturbations at the cosmic microwave background (CMB) scale.  

Another interesting prediction of our scenario is the formation of multi-quark and lepton nuggets \cite{Bai:2018vik}, carrying the conserved $B-L$ charge. We estimate the typical mass of such nuggets is $\sim 10^{9}\text{ kg}$, and demonstrate that they are long-lived to survive to present days. However, they contribute less than a per cent to the dark matter density  $\Omega_{\text{QN}}/\Omega_{\text{\rm dm}}\sim 6\times 10^{-3}$.

Finally, we examine the production of gravitational waves in our scenario. We find that gravitational waves with a peak frequency of approximately $\sim 10^{-5}$ Hz and a fractional energy density of around $\sim 10^{-16}$ are predominantly generated by sound waves in the primordial plasma during the phase transition. Unfortunately, none of the planned gravitational-wave detectors are sensitive enough to observe such a stochastic gravitational-wave background. 

We must emphasise in advance that our predictions are based on a rather simplified modelling of highly complex dynamics and should therefore be regarded only as order-of-magnitude estimates.

The rest of the paper is organised as follows. In the next section, we discuss in detail the QCD and the electroweak phase transitions within the SM with hidden scale invariance. The characteristics of the gravitational waves produced during the phase transition are given in Sec. \ref{gw}. In Sec. \ref{blackholes}, our predictions for primordial black holes are presented. In Sec. \ref{nuggets}, we discuss the physics of quark-lepton nuggets and estimate their abundance.  Finally, Sec. \ref{dis} is reserved for summary and further discussions.

\section{QCD and electroweak phase transitions in the Standard Model with hidden scale invariance}
\label{phasetransition}

In this section, we discuss the QCD and electroweak phase transitions within the minimal SM featuring spontaneously broken scale invariance. This framework introduces a light dilaton field alongside the usual SM fields. The following sections will explore the physical implications of these phase transitions. 

\subsection{SM with hidden scale invariance}

The framework of classical scale invariance offers an appealing approach to understanding the origin of mass scales. Owing to the presence of an intrinsic anomaly, classically scale-invariant theories undergo dimensional transmutation once radiative corrections are taken into account \cite{Coleman:1973jx}. This phenomenon arises in both strongly and weakly coupled theories and is perhaps connected to the non-perturbative vacuum structure of general relativity \cite{Dvali:2024dlb}. A key consequence is the emergence of a corresponding pseudo–Goldstone boson—the dilaton—in the particle spectrum of the theory.

In this work, we remain agnostic regarding the precise mechanism of spontaneous scale (or, more generally, conformal) symmetry breaking, which we assume to occur in the ultraviolet regime of the theory (see, e.g., \cite{Armillis:2013wya}). Furthermore, we take the effective field theory below the conformal symmetry–breaking scale to be the minimal SM.\footnote{This is a deliberately conservative assumption. Our framework, however, can be readily adapted to any beyond-the-Standard-Model scenario, which may yield quantitatively different low-energy predictions. Such cases should be analysed individually.}. 

Despite the unknown details of spontaneous conformal symmetry breaking, the dilaton enters the low-energy theory in a unique and well-defined manner through the nonlinear realisation of the broken symmetry. In practice, every classical parameter of the SM with a nonzero mass dimension—such as the electroweak Higgs vacuum expectation value (VEV), $v_{ew}$, the cosmological constant, $\lambda_c$, and the cutoff scale $\Lambda$ — is promoted to depend on the dynamical dilaton field. This procedure renders the effective SM manifestly scale-invariant. Respectively, one obtains the characteristic dilaton–Higgs coupling, $\sim \chi^2 |H|^2$, the dilaton self-interaction, $\sim \chi^4$, and the scale anomaly–induced coupling of the dilaton to the trace of the energy–momentum tensor, which arises upon expressing the dilaton-dependent effective running couplings:    
\begin{eqnarray} g_i(\chi)=g_{i}(\Lambda)+\beta_{g_i}\ln\left(\frac{\chi}{\Lambda} \right)+\frac{1}{2!} \beta'_{g_i}\ln^2\left(\frac{\chi}{\Lambda} \right)+...
    \label{lag}
\end{eqnarray}
In the above Taylor series expansion, $\beta_{g_i}=\frac{dg_i}{d\ln(\chi)}\sim \mathcal{O}(\hbar)$, is the renormalisation group (RG) beta-function for a dimensionless coupling $g_i$, while the coefficients of sub-leading $\ln$-terms are $\beta^{(n)}_{g_i}=\frac{d^{n+1}g_i}{d\ln(\chi)^{n+1}}\sim \mathcal{O}(\hbar^n)$. Without loss of generality, we define the cut-off scale as the VEV of the dilaton field, $\langle \chi \rangle \equiv v_{\chi}=\Lambda$. Neglecting infrared-irrelevant, higher-dimensional operators, the above procedure uniquely determines the Lagrangian of the theory. 

Let us now focus on the scalar potential of the model (at zero temperature, $T=0$), retaining only the neutral component of the electroweak doublet Higgs field, $H^T=(0,h/\sqrt{2})$ (unitary gauge fixing):
\begin{equation}
V_{0}(h,\chi) = \frac{\lambda_h(\chi)}{4}\left(h^2 - \xi(\chi)\chi^2\right)^2 + \frac{\lambda_\chi(\chi)}{4}\chi^4.
\label{pot}
\end{equation}
Note that this potential already incorporates radiative corrections through the effective running coupling constants, and that $h$ and $\chi$ are understood as the `classical' fields of the quantum-corrected effective potential.

Due to the underlying scale invariance, the shape of this potential differs significantly from the standard `Mexican hat' Higgs potential. Specifically, the potential is flat at the origin,
\begin{equation}
\left.\frac{d^2V_0}{dh^2} = \frac{d^2V_0}{d\chi^2} = \frac{d^2V_0}{dhd\chi}\right\vert_{h=\chi=0} = 0,
\label{flat}
\end{equation}
whereas the SM Higgs potential exhibits a large negative curvature already at tree level. Further insight arises from the phenomenological requirement of cancelling the large cosmological constant (i.e., the potential energy at its minimum). This implies that the potential at the origin and at its minimum are (nearly) degenerate,\footnote{Interestingly, this is precisely the requirement for the phenomenologically successful multiple-point criticality proposed in Ref.~\cite{Froggatt:1995rt}. It occurs automatically in scale-invariant theories~\cite{Foot:2014ifa}, where the cancellation of the vacuum energy contribution relies on the tuning of dimensionless couplings~\cite{Foot:2014ifa}, rather than by adding a `bare' cosmological constant. The latter approach, typically assumed in the literature, is incompatible with our scenario of non-linearly realised scale invariance.}
\begin{equation}
V(0,0) = V(v_{ew}, v_{\chi}) = 0.
\label{deg}
\end{equation}
This has important ramifications for the dilaton mass and, as will be shown subsequently, for the cosmological phase transition. 

Specifically, let us first analyse the extrema of the potential~(\ref{pot}). The solution to the extremum equation in the $h$-direction,
\begin{equation}
\left. \frac{\partial V_0}{\partial h}\right\vert_{h=v_{ew},~\chi=\Lambda}=
\lambda_h(\Lambda)v_{ew}\left(v_{ew}^2-\xi(\Lambda)\Lambda^2\right)=
0,
\label{extr1}
\end{equation}
\begin{equation}
\frac{v_{ew}^2}{\Lambda^2} = \xi(\Lambda)~,
\label{extr12}
\end{equation}
implies that the hierarchy between the electroweak scale $v_{ew}$ and the ultraviolet scale $\Lambda$ is determined exclusively by the dimensionless parameter $\xi$. This parameter receives only mild logarithmic radiative corrections due to the trace anomaly. Furthermore, $\xi=0$ is an attractive fixed point of the RG flow, leading to an enhanced emergent symmetry \cite{Berezhiani:2024rth} associated with the factorisation of the Hilbert spaces of the SM and dilaton sectors.\footnote{In view of the discussion in Ref. \cite{Dvali:2024dlb}, the full analysis must be incorporated within the supergravity picture. Here we simply assume that the supersymmetry is broken at a high-energy scale $\sim \Lambda$, and is irrelevant at low energies.} As a result, the large hierarchy of scales is technically natural \cite{tHooft:1979rat}; that is, radiative corrections do not spoil the tree-level hierarchy.

The extremum in the $\chi$-direction reads: 
\begin{equation}
\left. \frac{\partial V_0}{\partial \chi}\right\vert_{h=v_{ew},~\chi=\Lambda}
=\frac{\beta_{\lambda_{\chi}}}{4}\Lambda^4 =0~, 
\end{equation}
where we have used Eq. (\ref{extr12}) and $\lambda_{\chi} (\Lambda)=0$. The latter follows from Eq. (\ref{extr12}) and the condition of vanishing vacuum energy, Eq. (\ref{deg}). Thus, the extremum derivative in $\chi$-direction is defined by the running of the dilaton self-interaction coupling ($\sim \mathcal{O}(\hbar)$ effect), and the extremum equation implies:
\begin{equation}
\beta_{\lambda_{\chi}}(\Lambda)=0,~\text{alongside with}~\lambda_{\chi}(\Lambda)=0. 
    \label{}
\end{equation}
In other words, the tuning of the cosmological constant as described by the condition (\ref{deg}) is also technically natural in our scenario. Note that, solving for $\Lambda$, the first equation determines the overall scale of the theory through the dilaton VEV (\emph{the dimensional transmutation} \cite{Coleman:1973jx}), while the second equation ensures cancellation of the vacuum energy through the tuning of parameters involved implicitly in the effective running coupling $\lambda_{\chi}$ when evaluated at $\Lambda$.  

Next, we turn to the curvature of the potential to determine the scalar mass spectrum. At the extremum, we have:
\begin{eqnarray}
\frac{\partial^2 V_0}{\partial h^2}&=&2\lambda_h(\Lambda)v_{ew}^2, \\
\frac{\partial^2 V_0}{\partial \chi^2}&=&2\lambda_h(\Lambda)v_{ew}^2\left(1+\frac{\beta_{\xi}(\Lambda)}{2\xi(\Lambda)}\right) \nonumber \\
&+&\frac{1}{4}\beta'_{\lambda_{\chi}}(\Lambda)\frac{v_{ew}^2}{\xi^2(\Lambda)},\\
\frac{\partial^2 V_0}{\partial h\partial\chi}&=&-2\lambda_h(\Lambda)v_{ew}^2\left(1+\frac{\beta_{\xi}(\Lambda)}{2\xi(\Lambda)}\right)
\end{eqnarray}
It is quite remarkable that the determinant of the corresponding Hessian is proportional to
$\propto \beta'_{\lambda_{\chi}}\sim \mathcal{O}(\hbar^2)$. In other words, the potential maintains a flat direction at the one-loop level, and the dilaton acquires its mass only at the two-loop level. This is a generic feature of the scale-invariant theories with natural tuning of the cosmological constant, which was first observed in linearly realised scale-invariant models \cite{Foot:2010et}.

To ensure that the extrema identified above correspond to true minima of the potential, the parameters must be evolved from the cutoff scale down to the electroweak scale. The positivity of the Hessian at that scale should then be verified. This analysis was performed in \cite{Arunasalam:2017ajm} at the two-loop level. We quote below the results from that work for the Higgs and dilaton masses, as well as the mixing angle between these two fields:
\begin{eqnarray}
    m_h \simeq \sqrt{2\lambda_h(v_{ew})}v_{ew}\approx 125~\text{GeV} \label{mh}\\
    m_{\chi}\sim\frac{\xi(v_{ew})}{16\pi^2}v_{ew} \approx 10^{-8}~\text{eV}, \label{mchi}\\ 
    \tan\alpha \simeq -\xi \approx -10^{-16}, \label{angle}
\end{eqnarray}
for the dilaton VEV, $\Lambda\sim M_{P}=2.4\cdot 10^{18}$ GeV. 

We note that, despite the extension of the SM by the dilaton sector, the predictive power of the theory remains unchanged. This is a consequence of the absence of explicit mass parameters in the potential (\ref{pot}) and the imposed condition of vanishing vacuum energy (\ref{deg}), both of which arise from the consistent realisation of hidden scale invariance. Consequently, the dilaton mass and the mixing angle are fully determined in terms of the SM parameters, including an ultraviolet cutoff that is a part of the effective theory description. Recall, in our scenario, the cutoff is identified with the vacuum expectation value of the dilaton field that emerges as a result of dimensional transmutation. 

From Eqs. (\ref{mh}, \ref{mchi},\ref{angle}), we observe that both the dilaton mass and its mixing with the SM Higgs boson are proportional to the parameter $\xi$, which characterises the hierarchy between the electroweak and ultraviolet scales. Consequently, a theory that naturally accommodates a large hierarchy of scales in the low-energy domain contains, in addition to the SM fields, a light dilaton that couples only feebly to them. From the standpoint of collider phenomenology, such a theory is therefore nearly indistinguishable from the SM. However, as we will demonstrate in the remainder of this paper, this minimal scale-invariant framework gives rise to a drastically different scenario for the cosmological evolution of the Universe.

\subsection{Effective potential at finite temperature}
We now proceed to analyse our theory in a cosmological setting, beginning with its behaviour in the presence of a primordial plasma in thermal equilibrium at a temperature $T$. The thermal corrections to the zero-temperature scalar potential (\ref{pot}), prior to the hadronisation phase transition, are dominated by the heaviest Standard Model particles: the top quark, the Higgs boson, and the electroweak gauge bosons. In the high-temperature expansion, the thermal correction to the potential takes the form
\begin{eqnarray}
V_T &=& V_0 -\frac{\lambda_h \xi}{24} T^2 \chi^2 \nonumber \\
&+&\frac{1}{24}\left(3\lambda_h +3 y_t^2+\frac{9}{4}g^2+\frac{3}{4}g'{}^2\right)T^2 h^2.
\label{thermal1}
\end{eqnarray}
We immediately observe that finite-temperature effects explicitly break scale invariance, in addition to the breaking induced by quantum corrections. In the high-temperature regime, the former dominate, and we therefore consider temperature-dependent extrema of the classical scalar fields, which we denote by barred quantities in what follows. In particular, the thermal expectation value of the dilaton field is given by
\begin{equation}
\bar v_{\chi}=\frac{1}{\xi}\left(\bar{v}^2_{ew}+\frac{T^2}{12}\right).
\label{thermdil}
\end{equation}
As a consequence of the assumed hierarchy, $\xi \ll 1$, the dilaton acquires a large thermal expectation value, $\bar v_{\chi} \gg T$, and thus contributes to the dynamics of the phase transition predominantly through its thermal energy. We may therefore integrate out the dilaton field using Eq.~(\ref{thermdil}), and focus henceforth on the dynamics of the Higgs field.

Since the zero-temperature potential along the Higgs direction is flat at the origin—a direct consequence of the imposed scale symmetry—and the thermal correction in the second line of Eq.~(\ref{thermal1}) induces a positive curvature (with $\lambda_h>0$ required for stability), the Higgs field becomes trapped in a symmetric extremum at the origin, $\bar v_{ew}=0$. Consequently, no electroweak phase transition can occur prior to the QCD phase transition. This behaviour is a rather generic prediction of a class of scale-invariant theories with the basic features discussed in the previous section, and it extends naturally to models beyond the minimal SM.    

\subsection{Dynamics of the QCD triggered phase transition}

As discussed in the previous section, in our model, the electroweak phase transition does not occur until the Universe cools to temperatures comparable to those of the QCD phase transition. In fact, two distinct phase transitions take place. The first is the QCD confinement transition, during which quarks and gluons become confined into composite hadrons and mesons. This transition generates massive composite hadrons, as well as a massive meson associated with the anomalous axial symmetry, while the pions remain massless because the quarks are fundamentally massless. The confinement transition is followed by the QCD chiral symmetry–breaking phase transition, which in turn triggers electroweak symmetry breaking. Once these transitions are completed, the subsequent cosmological evolution proceeds as in the Standard Model. In what follows, we will use the superscript $(h)$ to identify quantities related to the hadronisation transition, while the superscript $(\chi)$ will refer to the QCD chiral phase transition. 

\subsection*{\it QCD hadronisation phase transition}

In most of the treatments of the problem, the QCD hadronisation and chiral phase transitions are often assumed to occur nearly simultaneously. In our scenario, however, the situation is qualitatively different. First, since all Standard Model particles remain massless prior to these transitions, the confinement phase transition and the associated formation of non-perturbative gluon condensates occur at significantly lower temperatures. This scale can be estimated by determining when QCD with six massless quarks enters the strongly coupled regime. We adopt the results of Ref.~\cite{ParticleDataGroup:2016lqr}, where this scale was computed to be $\Lambda^{(6)}_{\text QCD}\approx 89$ MeV. We therefore assume that the confinement phase transition occurs at around a critical temperature of $T_c^{(h)}\lesssim \Lambda_{\text QCD}$\cite{Braun:2006jd}.

More specifically, the critical temperature for the hadronisation phase transition is defined as a temperature where the pressure across the boundary of the old phase of quark-gluon plasma and new phase with hadronic matter is balanced. The later includes the pressure exerted by the gluon condensates, defined through the 'bag model' parameter $B\approx \Lambda^{(6)~4}_{QCD}$. Hence, we have: $\frac{\pi^2}{90}g_*^{(h)}T_c^{(h)~4}=\frac{\pi^2}{90}g_*^{(q)}T_c^{(h)~4}-B$, where $g_*^{(q)}\approx 106.75$ and $g_*^{(h)}=62.75$ are the effective relativistic degrees of freedom in the quark-gluon and hadronic phases, respectively. Solving the balance equation we find
\begin{equation}
T^{(h)}_{c}=\left(\frac{90}{\pi^2}\frac{B}{g_{*}^{(q)}-g_{*}^{(h)}}\right)^{1/4}\approx 60~\text{MeV}\,.
    \label{hcrit}
\end{equation}

The detailed modelling of this phase transition is complicated due to the non-perturbative dynamics involved. We just assume in this paper that the completion of the phase transition happens almost at the critical temperature. For further details, see Sec. \ref{nuggets}. 

\subsection*{\it QCD chiral phase transition}

Now we turn to the QCD chiral phase transition. Both analytical arguments \cite{Pisarski:1983ms} and lattice simulations \cite{Cuteri:2017gci} seem to indicate that QCD with six massless quarks undergoes a first-order phase transition, in sharp contrast to the standard case with three light quarks. The case for the first-order QCD phase transition is stronger within our scenario with the underlying scale invariance. This suggests that the chiral phase transition is delayed and occurs at a lower temperature than the confinement transition, rather than coinciding with it. To model the chiral-electroweak phase transition, we consider $SU(6)\times SU(6)$ symmetric linear sigma model coupled through the Higgs boson, together with the massless leptons and electroweak gauge bosons. At temperatures $T< T_c^{(h)}$, we ignore all massive composite states, hadrons, glueballs and alike. 

Keeping only the leading marginal operators, the scalar potential that describes pion-Higgs interactions takes the following form:\footnote{For simplicity, we ignore a quartic invariant, $\propto {\rm tr}\left(\Phi^{\dagger}\Phi\Phi^{\dagger}\Phi\right)$, that is subdominant for large $n_f$.}  
\begin{equation}
V_{\sigma-h} = \lambda_\sigma (\tr\Phi^\dagger\Phi)^2 -\lambda_\kappa \tr (\Phi^\dagger\Phi \Phi M) +h.c. 
\label{sigmapot}
\end{equation}
The first term describes the linear \(\sigma\)-model with 35 composite pion states, \(\pi^a\), and the \(\sigma\) meson:
\begin{equation}
\Phi = \frac{\sigma}{\sqrt{2n_f}}
\exp\!\left(\frac{i\sqrt{2n_f}\pi^a T^a}{v_\sigma}\right)\!,
\label{meson}
\end{equation}
where \(T^a\) are the \(SU(6)\) generators in the fundamental representation (\(a=1,2,\ldots,35\)), \(n_f=6\), and \(v_\sigma\) denotes the order parameter for the \(SU(6)\times SU(6)\) chiral symmetry breaking.  

The second term in Eq.~(\ref{sigmapot}) describes the interaction of the composite meson fields with the Higgs field. It originates from the quark--Higgs Yukawa couplings of the Standard Model in the unconfined regime and thus takes the form
\begin{equation}
M = 
\begin{pmatrix}
y_u&0&0&0&0&0\\
0&y_d&0&0&0&0\\
0&0&y_s&0&0&0\\
0&0&0&y_c&0&0\\
0&0&0&0&y_b&0\\
0&0&0&0&0&y_t
\end{pmatrix}
\frac{h_c}{\sqrt{2}}\,.
\end{equation}
This term explicitly violates the \(SU(6)\times SU(6)\) chiral symmetry and generates pion masses after spontaneous chiral symmetry breaking and electroweak symmetry breaking.

Importantly, this interaction plays a crucial role in the electroweak phase transition. Expanding around the chiral symmetry–breaking expectation value of the \(\sigma\) field produces a term linear in the Higgs field, which tilts the Higgs potential and drives the Higgs field toward the electroweak symmetry–breaking configuration.

A few comments are in order. First, owing to the underlying scale invariance, the potential~(\ref{sigmapot}) contains no explicit mass parameter. Consequently, symmetry breaking relies on radiative corrections and, as discussed above, the transition is of first order. Second, in Eq.~(\ref{sigmapot}) we have neglected the dilaton-meson coupling. This is justified in the regime where both the dilaton and the Higgs are trapped near the origin of the potential and their classical values are related through Eq.~(\ref{thermdil}). Using this relation would convert the dilaton-meson interaction into an effective Higgs-meson coupling. This effective coupling, however, is suppressed by a factor of order \(f_\pi/v_{\rm ew}\sim10^{-6}\) compared to the dominant top-quark–induced Higgs-meson coupling and can therefore be safely neglected. Once chiral and electroweak symmetry breaking are complete, the heavy quarks decouple, and the Higgs interactions with the light quarks generate the masses and decay constants of the standard light mesons observed today. Finally, in Eq.~(\ref{sigmapot}) we have neglected the higher-dimensional term \(\propto \det\Phi + {\rm h.c.}\), originating from the non-perturbative ’t~Hooft vertex. This term breaks the axial \(U(1)_A\) symmetry, while preserving $SU(6)\times SU(6)$ symmetry, and is responsible for the large \(\eta'\) meson mass, which we assume to be decoupled from thermal plasma at $T\lesssim T_c^{(h)}$.

To study the phase transition, it is convenient to project onto $\sigma-h$ direction in the scalar field space and introduce the polar representation of $(\sigma,h)$ fields, $(\rho, \theta)$, where $\rho=\left(\sigma^2+h^2\right)^{1/2}$ and $\theta=\arctan(\sigma/h)$. The tree-level Lagrangian in $\sigma-h$ direction then reads:
\begin{eqnarray}
V_0(h,\sigma) &=& 
\frac{\lambda_\sigma}{4}\sigma_c^4 - \frac{\lambda_\kappa y_t}{2 n_f^{3/2}}\sigma^3 h \nonumber \\
&=&\left(\frac{\lambda_\sigma}{4} \sin^4\theta - \frac{\lambda_\kappa y_t}{2 n_f^{3/2}}\sin^3\theta\cos\theta\right)\rho^4 
\label{pot0}
\end{eqnarray}
Because of the classical scale invariance, we observe that the modulus field $\rho$ enters (\ref{pot0}) as $\lambda_{\rho}\rho^4$. As discussed in Appendix \ref{linsigparameters}, the value of the effective dimensionless couplings $\lambda_{\sigma}$ and $\lambda_{\kappa}$ are estimated to be large, compromising the perturbative approximation. 
However, their $\theta$-dependent  combination, $\lambda_{\rho}$, is actually small, which allows to compute quantum and thermal corrections in the $\rho$-direction perturbatively. In fact, the effective coupling is  $\lambda_{\rho}\approx -0.3$ 
when evaluated at the minimum of $\theta$ in the tree-level approximation, 
\begin{equation}
    \tan\theta = -\frac{\lambda_{\sigma}n_f^{3/2}}{\lambda_{\kappa}y_t}\left(1-\sqrt{1+3\left(\frac{\lambda_{\kappa}y_t}{\lambda_{\sigma}n_f^{3/2}}\right)^2}\right) 
    \Longrightarrow \theta \approx 0.7\,.
    \label{theta}
\end{equation}
We verified numerically that this approximate tree-level expectation value is very close to the value obtained by minimising the full one-loop potential. Hence, the one-loop quantum,  
\begin{equation}
\delta V = \sum_i g_i \frac{m_i^4(T=0)}{64\pi^2}\log \left(\frac{m_{i}^2(T=0)}{\Lambda_{QCD}^2}\right)\, \end{equation}
and finite temperature, 
\begin{equation}
\delta V_T(\rho,\theta,T) = \sum_i g_i\frac{T^4}{2\pi^2} J_B\left(\frac{m_{i}^2}{T^2}\right)\,,
\end{equation}
corrections can be expressed through the field-dependent masses given in Table \ref{table1}. We included only dominant contributions to the full effective potential, $V_{eff}=V_0+\delta V+\delta V_T$, as well as one-loop thermal contribution to masses to effectively account for the resummation of infrared daisy diagrams \cite{Arnold:1992rz}.
\begin{table}[t]
\begin{center}
\renewcommand{\arraystretch}{1.4}
    \begin{tabular}[t]{|c|c|c|c|}
        \hline  
        Particle & Field dependent mass$^2$ $m_i^2(\rho,\theta,T)$ & \# of relativistic dofs, $g_i$ \\ \hline  
        $W_t$ & $\frac{1}{4}g^2\rho^2\cos^2\theta$ & 4  \\
        $W_l$ & $\frac{1}{4}g^2\rho^2\cos^2\theta +\frac{11}{6}g^2T^2$ & 2  \\
        $Z_t$ & $\frac{1}{4}(g^2+g'{}^2)\rho^2\cos^2\theta$ & 2  \\
        $Z_l$ & $\frac{1}{2}\left(\frac{1}{4}(g^2+g'{}^2)\rho^2\cos^2\theta+\frac{11}{6}(g^2+g'{}^2)T^2+\Delta(\rho,\theta,T)\right)^*$ & 1  \\
        $\gamma_l$ & $\frac{1}{2}\left(\frac{1}{4}(g^2+g'{}^2)\rho^2\cos^2\theta+\frac{11}{6}(g^2+g'{}^2)T^2-\Delta(\rho,\theta,T)\right)^*$ & 1  \\
        $\pi_{35}$ & $\frac{5}{3}\frac{\lambda_\kappa y_t}{4 n_f^{1/2}}\rho^2\sin\theta\cos\theta $ & 1  \\
        $\pi_{25-34}$ & $\frac{\lambda_\kappa y_t}{4 n_f^{1/2}}\rho^2\sin\theta\cos\theta $ & 10  \\
        \hline      
    \end{tabular}
    \end{center}
    \caption{The effective relativistic degrees of freedom and their field-dependent masses that give the dominant contribution to the one-loop effective potential $V_{eff}$ at $T\lesssim T_c^{(h)}\approx 60$ MeV. This contribution is dominated by pion states spanning the coset $SU(6)\times SU(6)/SU(6)$ of spontaneously broken chiral flavour symmetry; 
    $\Delta(\rho,\theta,T) = \sqrt{\left(\frac{1}{4}(g^2+g'{}^2)\rho^2\cos^2\theta+\frac{11}{6}(g^2+g'{}^2)T^2\right)^2-g^2g'{}^2\frac{11}{3}T^2(\frac{11}{3}T^2+\rho^2\cos^2\theta)}$.}
    \label{table1}
\end{table}

At temperatures $T\lesssim T_c^{(h)}\approx 60$ MeV, the full effective potential is dominated by the thermal corrections and develops a second minimum away from the one at the origin, $\rho=0$. We found that two minima become degenerate at $T_c^{(\chi)}\approx 28$ MeV. This is the onset of the chiral-electroweak phase transition that proceeds via the nucleation of bubbles with non-zero expectation value of $\rho$. 

We distinguish two regimes in which nucleation of $\rho$-bubbles of the new phase may occur. One is the production of bubbles with a typical size smaller than a thermal cell. This process is dominated by quantum tunnelling, where the thermal effects are negligible. In the semiclassical approximation, the probability for nucleation of large bubbles with size $R>1/T$ is described by $O(4)$-symmetric classical solutions of the $\rho$-field. Since $\lambda_{\rho}<0$ at the classical level, the relevant solutions are given by the scale-invariant Fubini--Lipatov instantons~\cite{Fubini:1976jm}. However, we verified numerically that, at $T\sim T_c^{(\chi)}$, production of small bubbles via quantum fluctuations is highly improbable.

The nucleation rate for large bubbles, $R>1/T$, proceeds through thermal fluctuations, instead.  It is defined by $O(3)$ symmetric bounce solutions of the following static (in Euclidean time) equation, 
\begin{equation}
\frac{d^2 \rho}{dr^2}+\frac{2}{r}\frac{d \rho}{dr}-\frac{\partial V_{\text{eff}}}{\partial \rho}=0\,,
\end{equation}
with boundary conditions $\frac{d \rho}{dr}\big|_{r=0}=0$ and $\lim_{r\rightarrow\infty} \rho(r)=0$. We employ the standard shooting method to numerically solve this equation. The solution then is used to compute the bubble nucleation rate at fixed temperature $T$, 
\begin{equation}
\Gamma (T)= A(T){\rm e}^{-S(T)}\,,   
\end{equation}
where the thermal action and the prefactor are defined as:
\begin{equation}S[\rho,T] = \frac{4\pi}{T}\int_0^\infty dr\, r^2\left(\frac{1}{2}\left(\frac{d\rho}{dr}\right)^2+V_{\text{eff}}(\rho,T)-V_{\text{eff}}(0,T)\right)\,
\end{equation}
\begin{equation}
A(T) = T^4 \left(\frac{S}{2\pi}\right)^{3/2}\,.
\end{equation}

The produced bubbles of the new phase subsequently grow, collide, and percolate, such that the new phase eventually occupies a large fraction of the Hubble volume, after which the phase transition is complete. To determine when the phase transition completes, we compute the probability for a given point within the Hubble volume to remain in the old $SU(6)\times SU(6)$ symmetric phase at time $t$, expressed in terms of the expansion scale factor $a(t)$:
\begin{equation}
p(x) = \exp\left(\frac{-3}{16\pi \left(\frac{\pi^2}{30}g_*^{(h)}\right)^2(2\pi)^{3/2}}\frac{m_P^4}{T_c^{(\chi)}{}^4}
\int_{1}^x dx' \,(x-x')^3
S\!\left(\frac{T_c^{(\chi)}}{x'}\right)^{3/2}
e^{-S\!\left(\frac{T_c^{(\chi)}}{x'}\right)}\right)\,,
\label{oerc}
\end{equation}
where $x=a/a_c$, with $a_c$ denoting the scale factor at the critical temperature $T_c^{(\chi)}\simeq 28~\mathrm{MeV}$, and $g_*^{(h)}=63.75$ is the effective number of relativistic degrees of freedom in the symmetric phase at $T_c^{(\chi)}$. In deriving the probability~(\ref{oerc}), we assume that the Universe remains radiation dominated throughout the phase transition. This assumption is well justified in our case, since the vacuum energy density constitutes a subdominant contribution to the total energy budget due to the approximate scale invariance of the underlying theory.

According to percolation theory~\cite{Shante:1971}, the critical fraction of a Hubble volume that must be occupied by the new phase to ensure completion of the phase transition is approximately $1/3$. We therefore set $p\simeq 0.7$ and obtain the percolation temperature $T_p^{(\chi)}\simeq 26.5~\mathrm{MeV}$, which lies close to the critical temperature, $T_c^{(\chi)}$.

Another important parameter in our analysis is the velocity of the expanding bubbles of the new phase. A detailed study, presented in Appendix \ref{bubblespeed}, shows that the bubble walls become highly relativistic, $v_w\simeq 1$, although they never enter the runaway regime.

To summarise this section, the confinement phase transition in our scenario occurs at $T_c^{(h)}\simeq 60~\mathrm{MeV}$, while the system remains in the electroweak-symmetric phase. Substantial further cooling of the Universe to $T_c^{(\chi)}\simeq 28~\mathrm{MeV}$ is required to trigger the chiral phase transition, which completes at $T_p^{(\chi)}\simeq 26.5~\mathrm{MeV}$ through nucleation, expansion, and percolation of relativistic bubbles associated with the breaking of $SU(6)\times SU(6)\to SU(6)$ chiral symmetry. The electroweak phase transition is then induced by the QCD chiral transition and completes almost instantaneously via classical rolling of the Higgs field toward its electroweak-breaking minimum. A number of interesting physical phenomena may occur during this sequence of transitions, to which we now turn.

\section{Gravitational wave spectrum}
\label{gw}
Having established the details of the QCD chiral phase transition, we are now ready to investigate the production of gravitational waves.  First-order phase transitions generate gravitational waves through three principal mechanisms: collisions of expanding bubble walls, sound waves propagating through the plasma within the bubbles, and magnetohydrodynamic turbulence in the plasma \cite{Caprini:2015zlo}. The dominant contribution depends on how the released vacuum energy is partitioned among these channels.

In runaway transitions, the fraction of energy transferred to the plasma saturates, leaving the remaining energy to be deposited into the bubble walls. As a result, the bubble walls continue to accelerate, and gravitational waves from bubble wall collisions can become the dominant contribution. However, as demonstrated in Appendix \ref{bubblespeed}, the walls separating chiral symmetric and asymmetric phases never enter the runaway regime. For non-runaway transitions, the situation is different. The energy stored in the bubble walls scales with the surface area of the expanding bubble, whereas the energy transferred to the surrounding plasma scales with the bubble volume. Consequently, the plasma energy fraction rapidly exceeds the wall contribution, making sound waves and magnetohydrodynamic turbulence the dominant sources of gravitational-wave production.

Gravitational waves produced in first-order phase transitions have been modelled by coupled scalar-hydrodynamic simulations on a lattice \cite{Hindmarsh:2015qta},  the results of which are fit reasonably well by the envelop approximation \cite{Caprini:2015zlo}, which gives the gravitational wave spectrum as a function of a few key phase transition parameters. These include: the ratio between the phase transition inverse duration, $\beta \equiv -\frac{dS}{dt}\big|_{t=t_p}$, and the Hubble rate at this time, which is given by
\begin{equation}
    \frac{\beta}{H} = T_p^{(\chi)}\frac{dS}{dT}\big|_{T=T_p^{(\chi)}}\approx 4200,
\end{equation}
the Hubble rate redshifted to today,
\begin{equation}
    h=1.7 \times 10^{-5}\, \text{Hz}\,\left(\frac{T_p^{(\chi)}}{100  \,\text{GeV}}\right)\left(\frac{g_*}{100}\right)^{1/6},
\end{equation}
the ratio between the vacuum energy released by the transition and the surrounding radiation energy,
\begin{equation}
    \alpha = \frac{\epsilon_{\text{vac}}}{\epsilon_{\text{rad}}} \approx0.06,
\end{equation}
and the fraction of that released energy that goes into the bulk motion of the fluid, 
\begin{equation}
    \kappa_v = \frac{\alpha}{0.73+0.083\sqrt{\alpha}+\alpha}\approx 0.07.
\end{equation}
For gravitational waves produced from sound waves, the envelope approximation gives the gravitational wave energy density spectrum, redshifted to today, as 
\begin{equation}
    \Omega_{\text{SW}}(f)h^2 = 2.65 \times 10^{-6} \,\left(\frac{H}{\beta}\right) \left(\frac{\kappa_v \alpha}{1+\alpha}\right)^2 \left(\frac{100}{g_*}\right)^{1/3} v_w\, S_{\text{SW}}(f) \left( (8\pi)^{1/3}v_w\left(\frac{H}{\beta}\right)\left(\frac{\kappa_v \alpha}{1+\alpha}\right)^{-1/2}\right),
\end{equation}
where the factor $\left((8\pi)^{1/3}v_w\left(\frac{H}{\beta}\right)\left(\frac{\kappa_v \alpha}{1+\alpha}\right)^{-1/2}\right)$ accounts for the fact that the source in this case lasts for a shorter duration than the Hubble time \cite{Caprini:2015zlo}. The envelope determining the shape of the spectrum is given by
\begin{equation}
    S_{\text{SW}}(f) = \left(\frac{f}{f_{\text{SW}}}\right)^3\left(\frac{7}{4+3(f/f_{\text{SW}})^2}\right)^{7/2},
\end{equation}
with a peak frequency of 
\begin{equation}
    f_{\text{SW}} = 1.9 \times 10^{-5}\, \text{Hz}\,\frac{1}{v_w}\left(\frac{\beta}{H}\right)\left(\frac{T_p^{(\chi)}}{10^5  \,\text{MeV}}\right)\left(\frac{g_*}{100}\right)^{1/6}.
\end{equation}
For the chiral-electroweak transition in this model, $f_{\text{SW}} \approx 2.0 \times 10^{-5}$ Hz with a peak amplitude $\Omega_{\text{SW}}(f_{\text{SW}})h^2 \approx 1.2 \times 10^{-16}$.\\

For gravitational waves produced by turbulence, the approximation gives the energy density spectrum
\begin{equation}
    \Omega_{\text{turb}}(f)h^2 = 3.35 \times 10^{-4} \,\left(\frac{H}{\beta}\right) \left(\frac{\kappa_{\text{turb}} \alpha}{1+\alpha}\right)^{3/2} \left(\frac{100}{g_*}\right)^{1/3} v_w\, S_{\text{turb}}(f) 
\end{equation}
with the envelope
\begin{equation}
    S_{\text{turb}}(f) = \left(\frac{f}{f_{\text{turb}}}\right)^3\left(\frac{1}{(1+f/f_{\text{turb}})^{11/3}(1+8\pi f/h_*)}\right)
\end{equation}
and the frequency defining the spectral shape (which is approximately the peak frequency)
\begin{equation}
    f_{\text{turb}} = 2.7 \times 10^{-5}\, \text{Hz}\,\frac{1}{v_w}\left(\frac{\beta}{H}\right)\left(\frac{T_p^{(\chi)}}{10^5  ~\mathrm{MeV}}\right)\left(\frac{g_*}{100}\right)^{1/6} 
\end{equation}
For this model, $f_{\text{turb}} \approx 2.8 \times 10^{-5}$ Hz with a peak amplitude $\Omega_{\text{turb}}(f_{\text{turb}})h^2 \approx 1.3 \times 10^{-19}$ which is far smaller due to the extra dependence on $h$. The contribution to the gravitational wave background from both sources lies far outside the sensitivity of all current and planned future gravitational wave detectors \cite{Renzini:2022alw}.

\section{Primordial black holes}
\label{blackholes}
The production of black holes in a primordial plasma requires the generation of local overdensities that are susceptible to subsequent gravitational collapse. One can envisage two distinct mechanisms for primordial black hole (PBH) production in our scenario. One is when the primordial overdensities are generated during the first-order phase transition, which occurs through the nucleation and expansion of the vacuum bubbles. Through this process, different patches of the universe may transition to a new phase at slightly different times. A patch that transitioned to a new phase later than its surroundings is in the false vacuum state and hence carries larger energy density. Another source of overdensities is quantum fluctuations during inflationary expansion of the universe\footnote{The inflation is naturally realised in scale-invariant scenarios like the one we are discussing here. See, e.g., \cite{Barrie:2016rnv}.}. Such overdensities are 'frozen' at super-horizon scales and undergo gravitational collapse when they enter the horizon.

The first-order phase transitions may provide favourable conditions for the gravitational collapse of overdensities in a primordial plasma. As described in Ref. \cite{Jedamzik:1996mr}, during a transition, a region that includes a mixture of high and low energy density phases has a reduced pressure response to compression. As the plasma is compressed, regions of the low-energy phase can convert back to the high-energy phase while staying at a constant pressure and temperature. This results in a lower critical overdensity required for the formation of a black hole and so increases the production of black holes at that time. 

The typical mass of such black holes, formed during the radiation era, is on the order of the radiation energy contained within a Hubble sound horizon at that time  
\begin{equation}
    m_{\text{PBH}} = \frac{4\pi}{3}(c_s H^{-1})^3\epsilon = \frac{1}{2}c_s\left(\frac{8\pi^3}{90}g_*^{(i)}\right)^{-1/2} \frac{m_p^3}{T_c^{(i)}{}^2}
\end{equation}
which depends on the relativistic degrees of freedom of the new phase and temperature at the time of the transition. For the confining transition occurring at around $T_c^{(h)}\simeq 60~\mathrm{MeV}$, with $g_*^{(h)} = 63.75$, the typical PBH mass is $m_{\text{PBH}}\sim  3M_\odot$, while for the chiral-electroweak phase transition occurring at around $T_c^{(\chi)}\sim 28~\mathrm{MeV}$, with $g_*^{(\chi)} = 10.75$, it is $m_{\text{PBH}}\sim  40 M_\odot$.

Now, we focus on the two mechanisms of the generation of overdensities in our model.  

\subsection{Primordial black holes formation due to the overdensities produced during the phase transition}
One possible source for the primordial overdensities come from the first-order phase transitions themselves \cite{Gouttenoire:2023naa}. A patch of the universe that transitions later than its surroundings will have a higher radiation density when it does transition due to the expansion of the universe. Radiation energy density reduces with the expansion, while vacuum energy does not, hence a late transitioning patch with its non-zero vacuum energy has a higher energy density than the radiation-filled patches around it. 

The overdensity for a late-transitioning patch is given by 
\begin{equation}
\label{overdensity}
    \delta(t_{\text{late}},t) \equiv \frac{\epsilon_{\text{rad}}^{\text{late}}(t_{\text{late}},t)-\epsilon_{\text{rad}}(t)}{\epsilon_{\text{rad}}(t)}.
\end{equation}
Simulations suggest that the critical overdensity required for collapse is $\delta_c\approx0.4-0.66$ \cite{Musco:2004ak}, though, as mentioned above, this may be lessened during a first-order phase transition. The numerator of Eq. (\ref{overdensity}) is expected to be less than the vacuum energy released by the transition, so to form a PBH, the late transition time $t_\text{late}$, must be after the time when $\frac{\epsilon_\text{vac}}{\epsilon_\text{rad}(t_\text{late})}=\delta_c$. Considering the chiral-electroweak transition, for $\delta_c = 0.4$, this corresponds to a false vacuum temperature of around $T\sim 16~\mathrm{MeV}$. Using Eq. (\ref{oerc}), we find that the probability of producing such primordial black holes is negligible. 

As emphasised in \cite{Flores:2024lng}, the PBH production due to overdensities (\ref{overdensity}) within the Hubble patch requires, in addition, the generation of curvature perturbations in order to ensure that the local overdensity is decoupled from the global Hubble expansion. If that is not the case, the more likely scenario for black hole production is due to the gravitational collapse of overdensities produced across the boundary between two phases. For this to happen, the width of the boundary and the gradient energy carried by the boundary wall satisfy the Schwarzschild criterion \cite{Flores:2024lng,Blau:1986cw}. Given the estimate of the typical width of domain walls and their energy, we find that this scenario is not applicable to our model as well. Therefore, the only feasible mechanism for PBH production during the phase transition in our model is due to the collapse of pre-existing overdensities.

\subsection{Primordial black holes from pre-existing inflationary overdensities}
Inflation is thought to produce primordial overdensities and spacetime curvature, which may be sufficient for significant PBH formation \cite{Ozsoy:2023ryl}. Typical inflation models predict over- and underdensities of a given scale, $k$, with a Gaussian distribution 
\begin{equation}
    f(\delta,k) = \frac{1}{\sqrt{2\pi}}\frac{1}{\sigma(k)}\exp\left(-\frac{1}{2}\frac{\delta^2}{\sigma(k)^2}\right)
\end{equation}
where the variance is given by 
\begin{equation}
    \sigma(M)\approx 5 \times 10^{-6}\left(\frac{M(k)}{5 \times 10^{23} M_{\odot}h^{-1}}\right)^{\frac{1-n}{6}},
\end{equation}
$M$ being the mass of pressure-less matter within one horizon corresponding to the scale $k$, and in the simplest inflation models, the spectral index $n=1$, which gives a scale-invariant spectrum. The fraction of horizon volumes at the time of the phase transition with the required overdensity to collapse to a black hole is then 
\begin{equation}
    \epsilon = \int_{\delta_c'}^{\infty}f(\delta,M)d\delta
\end{equation}
where $M\propto k^{-3}$ is the horizon mass corresponding to the scale $k$. By comparing to the radiation energy density at that time and appropriately scaling the energy density fraction of the black holes to today, one gets 
\begin{equation}
    \Omega_{\text{PBH}}h^2 = 5.8 \times 10^7 \left(\frac{T_{c}^{(i)}}{100~\mathrm{MeV}}\right)\left(\frac{g_*^{(i)}}{10.75}\right) \, \epsilon.
\end{equation}
Now, the overdensities at the horizon scale at the time of recombination are imprinted on the CMB, $\sigma_{\text{CMB}} = 5\times 10^{-6}$ \cite{COBE:1992syq}. If a scale-invariant spectrum is assumed, then the critical overdensity would need to be on the order of $\delta_c \sim 10^{-5}$ in order to get a non-negligible production of black holes. A blue-tilted spectrum with $n>1$ allows for non-negligible production at higher values for $\delta_c$ but is constrained by observed CMB anisotropies \cite{Bennett:1994gg} and spectral distortions \cite{Hu:1994bz}. Considering PBH formation during the hadronisation transition, if we generously allow $n\sim 1.62 - 1.63$ and $\delta_c\sim 0.05$, PBH densities of $\Omega_{\text{PBH}}h^2\sim 0.05 - 1$ can be achieved. If PBHs are produced during the chiral-electroweak phase transition then $\Omega_{\text{PBH}}h^2\sim 10^{-7}-10^{-4}$, for the same spectral tilt and critical overdensity. We observe that the quoted results depend exponentially on the spectral tilt and hence the details of the inflationary model. 

In summary, our model predicts production of PBHs during the QCD hadronisation or chiral phase transitions. They likely originate from the gravitational collapse of pre-existing overdensities. In the most favourable scenario, the typical mass and abundance of PBHs produced during the hadronisation phase transition are, $m_{\rm PBH}\sim 3M_{\odot}$ and $\Omega_{\text{PBH}}h^2\sim 0.05 - 1$, while PBHs produced during the chiral-electroweak phase transition have typical mass in the LIGO mass range, $m_{\rm PBH}\sim 40M_{\odot}$, and abundance $\Omega_{\text{PBH}}h^2\sim 10^{-7}-10^{-4}$. Hence, depending on the inflationary model, PBHs in our scenario may constitute a notable fraction of dark matter.

\section{Quark-Lepton nuggets}
\label{nuggets}
We now focus solely on the QCD hadronisation phase transition. As described above, due to the delayed electroweak symmetry breaking, six flavours of quark remain massless at this time, resulting in a first-order phase transition occurring at lower temperatures, $T_c^{(h)}\simeq 60~\mathrm{MeV}$, than in the Standard Model. Here we outline a potential model for how the confinement transition may proceed, proposed by ref. \cite{Kajantie:1986hq}, and its implications for the formation of the non-topological solitons that we call quark-lepton nuggets, which are similar to the solitons described in ref. \cite{Bai:2018vik}.

\subsection{Formation of quark-lepton nuggets during the hadronisation phase transition}
The scenario involves some small amount of supercooling of the universe beneath the critical temperature $T_c^{(h)}$, with the nucleation of bubbles of the hadron phase occurring at some lower temperature $T_n^{(h)}$. It assumes these bubbles expand sub-sonically as deflagrations and are preceded by shock fronts which travel at around the speed of sound in the plasma, $v_{\text{shock}}\sim c_s=1/\sqrt{3}$, and that these shock fronts carry off the majority of the energy from the transition. When shock fronts from neighbouring bubbles collide, their energy is converted into heat due to turbulence, which reheats the universe to some degree. If this reheating is sufficient to raise the temperature to or above $T_c^{(h)}$, then no further nucleation of bubbles can occur. The transition then proceeds only through the expansion of the preexisting bubbles, which occurs at constant pressure and temperature as the universe continues to expand. 

To model this scenario, we use the bag model with a bag constant given by the QCD confinement scale with 6 massless quarks, $B= (\Lambda_\text{QCD}^{(6)})^4=(89~\mathrm{MeV})^4$. At temperature $T_n^{(h)}$, the energy required to heat a radiation-dominated, mixed-phase patch of the universe (with quark phase volume $V_q$ and hadron phase volume $V_h$) back to $T_c^{(h)}$ is 
\begin{equation}
    E_{\text{required}} = V_q\frac{\pi^2}{30}g_*^{(q)}(T_c^{(h)}{}^4-T_n^{(h)}{}^4) + V_h\frac{\pi^2}{30}g_*^{(h)}(T_c^{(h)}{}^4-T_n^{(h)}{}^4)
\end{equation}
while the total energy released at this time is simply the difference in energy densities between the two phases multiplied by the volume of the hadron phase
\begin{equation}
    E_{\text{released}} = V_h \left(\frac{\pi^2}{30}(g_*^{(q)}-g_*^{(h)})T_n^{(h)}{}^4 +B\right).
\end{equation}
where we assume that the reheating after nucleation occurs faster than any further cooling from expansion can occur. By requiring that $E_{\text{released}}\geq E_{\text{required}}$, substituting the bag parameter for the critical temperature $B=\frac{\pi^2}{90}(g_*^{(q)}-g_*^{(h)})T_c^{(h)}{}^4$, and dividing by the full volume of the patch $V= V_q+V_h$, we get a requirement relating the volume fraction of the patch in the hadron phase when reheating occurs, $f_h$ and the degree of supercooling, $\frac{T_n^{(h)}}{T_c^{(h)}}$,
\begin{equation}
    f_h\geq \frac{3}{4} \frac{g_*^{(q)}}{g_*^{(q)}-g_*^{(h)}} \left(1-\frac{T_n^{(h)}{}^4}{T_c^{(h)}{}^4}\right).
\end{equation}
We stress that both $f_h$ and $\frac{T_n^{(h)}}{T_c^{(h)}}$ are unknown parameters of the transition. If the above requirement is satisfied, the transition will slow down dramatically and proceed on the timescale of expansion, $H^{-1}$.

If the universe is reheated and the nucleation of new bubbles is halted, then the existing bubbles will have some average distance between them, introducing a new characteristic length scale into the system. Importantly, at the time when the bubbles percolate, the quark phase regions in between the bubbles will have average sizes of order of this length scale. The scale can be computed by considering the bubble nucleation rate and the speed of the shock fronts \cite{Kajantie:1986hq}. For bubbles nucleated at a rate 
\begin{equation}
    p(T) = p_0 T_c^{(h)}{}^4 \exp\left(\frac{-w_0}{(1-T^4/T_c^{(h)}{}^4)^2}\right),
\end{equation}
 $w_0$ and $p_0$ being unknown parameters of the transition assumed to be order 1, the characteristic scale is given by 
 \begin{equation}
    R_{\text{sep}} = v_{sh}\frac{3}{2L}\left(\frac{w_0}{16L}\right)^{1/2}\left(\frac{g_*^{(q)}-g_*^{(h)}}{4g_*^{(q)}-g_*^{(h)}}\right)^{1/2}\frac{1}{\chi}
\end{equation}
with $L=\log\left(p_0 v_{sh}^3 (T_c^{(h)}/\chi)^4\right)$ and $\chi= \sqrt{\frac{8\pi}{3M_p^2}B}$ is the natural energy scale for the transition.

Provided the transition can continue to provide sufficient heat, the universe remains at the critical temperature and the bubble walls expand slowly, allowing the two phases to remain in thermal and chemical equilibrium. This period ends towards the end of the transition, after percolation, when the shrinking pockets of the quark phase are no longer able to supply the required heat. At this point, the temperature drops beneath the critical temperature and the unbalanced pressure forces on the transition walls will cause them to accelerate.

On one side of the transition wall, baryon number exists as massless quarks, while on the other, it is confined into heavy baryons. As a result, quarks experience a force at the transition wall, and a pre-existing baryon number asymmetry can become concentrated as the quark phase pockets rapidly shrink. Following the work done in \cite{Bai:2018vik}, we can very roughly estimate the amount of trapped charge in this scenario by assuming that chemical equilibrium is lost at the time of percolation and that the hadron phase baryon number density is consistent with the baryon asymmetry today. In the hadron phase, which will eventually fill the universe, the density at that time is given by $n_b^{(h)} = Y_b s(T_c^{(h)})$, where $Y_b\approx10^{-10}$ and the entropy density is $s = (2\pi^2/45)g_*^{(h)} T_c^{(h)}{}^3$. The assumption of equilibrium (up until percolation) then lets us relate the baryon number density in the two phases, $n^{(q)}_B = \frac{1}{r}n^{(h)}_B$, where the ratio $r\sim 5.1$ is derived in Appendix \ref{densityratio}. The total trapped charge in a quark phase pocket is then 
\begin{equation}
    Q_B = \frac{4\pi}{3}R_\text{sep}^3 n^{(q)}_B \sim 10^{36}
\end{equation}

\subsection{Stability of the quark-lepton nugget}
Due to the non-zero baryon number inside, quarks within the quark phase pockets cannot fully self-annihilate; hence, the shrinking pockets will not be able to collapse to nothing. Instead, the thermal and vacuum pressures on the phase transition walls will eventually be balanced by the fermion degeneracy pressure of the quarks trapped inside, forming a non-topological soliton.

This is not the full story, however. Due to the delayed electroweak transition, sphaleron processes remain unsuppressed at this stage in the universe's evolution, which allows for the conversion of baryon number into anti-lepton number, that is, quarks into anti-leptons. Thus, the relevant conserved charge for what will become the quark-lepton nugget is $B-L$ \footnote{In reality, there are three separately conserved global charges, $\frac{B}{3}-L_i$, one for each generation. The fermions are all degenerate in mass and their interaction rates, while different, are all far smaller than the relevant cooling time scale. Thus, we don't expect any relevant differences between the generations and can simply work with the total $B-L$.}. 

To model the quark-lepton nugget, we consider some total volume $V$ containing some spherical region of radius $R$, within which is a quark-phase plasma at temperature $T$ and vacuum energy density $B$ and surrounded by a hadron-phase plasma at the same temperature. Within the nugget, there is some net conserved $B-L$ number, which, for now, we assume cannot leak out of the nugget. With fixed temperature and volume, the relevant thermodynamic potential to minimise is the free energy with the added constraint of conserved $B-L$ number. The full definition of the free energy and its minimisation with respect to the $B-L$ chemical potential and the radius of the nugget can be found in Appendix \ref{free_energy}. After minimisation, we are left with the constrained free energy of the system,
\begin{equation}
    \Omega  = - P(T) V  +\mu_\text{B-L}(T) Q_\text{B-L} \\
\end{equation}
where $P(T) = \frac{\pi^2}{90}g_*^{(h)}T^4$ is the pressure throughout the system, balanced on either side of the phase transition walls, and the chemical potential is 
\begin{equation}
    \mu_\text{B-L} = \sqrt{-\beta +\sqrt{\beta^2+\gamma}}
\end{equation}
with $\beta = \frac{117\pi^2}{85}T^2$ and $\gamma = \frac{216\pi^2}{85}(B-\frac{\pi^2}{90}(g_*^{(q)}-g_*^{(h)})T^4)$. Inputting the parameters of the model, we find that the chemical potential is zero at the critical temperature and increases to $\sim 200~\mathrm{MeV}$ at $T=0$.

The above constrained free energy was derived by assuming that all the charge was trapped within the nugget and could not leak out. We are now interested in the change of this free energy as a unit of $B-L$ charge is moved from inside the nugget to the outside. Due to the delayed electroweak transition, all leptons are massless outside the nugget. Baryons, however, are massive due to confinement. Their masses can be estimated in six flavour QCD by rescaling the the three-flavour Standard model mass with the new confinement scale $m_B^{(6)} \sim m_B^{(SM)}\left(\Lambda_\text{QCD}^{(6)}/\Lambda_\text{QCD}^{(SM)}\right) = 938 \frac{89}{332} ~\mathrm{MeV} \sim 250~\mathrm{MeV}$.
Moving a unit of $B-L$ number from inside the nugget to outside will reduce the constrained free energy by $\mu_\text{B-L}(T)<200~\mathrm{MeV}$ and increase it by the mass of that unit of charge outside the nugget. Given that $m_B^{(6)}> \mu_\text{B-L}(T)$, the nugget will be unstable against the emission of leptons but not baryons.

After its formation, the nugget will begin to emit lepton number freely at least until the electroweak phase transition when the vacuum structure changes again. Whether the nugget evaporates before then will depend on the rates of various processes contributing to the leakage of $B-L$ number. First, we have the time for the universe to cool from the confining transition to the chiral-electroweak transition. During radiation domination, the scale factor evolves as 
\begin{equation}
    \frac{\dot{a}}{a} = H = \sqrt{\frac{8\pi^3}{90}g_*}\frac{T^2}{M_P}
\end{equation}
which, when integrated, gives
\begin{equation}
    t_f-t_i = \frac{1}{2}\sqrt{\frac{90}{8\pi^3 g_*}}M_P\left(\frac{1}{T_f^2}-\frac{1}{T_i^2}\right)
\end{equation}
To cool from $T_i=T_c^{(h)}\simeq60~\mathrm{MeV}$ to $T_f=T_c^{(\chi)}\simeq28~\mathrm{MeV}$ with $g_*^{(h)}=62.75$ takes $\Delta t\sim 6\times 10^{17}~\mathrm{MeV}^{-1}$.

Next is the rate of equilibration for sphaleron processes. The rate of change of baryon number density is 
\begin{equation}
    \frac{d n_B}{dt} = -n_g^2 \rho(0) \frac{\Gamma_{\text{diff}}}{T^3} (n_B-n_B^{eq})
\end{equation}
where $n_g^2 \rho(0)\approx 10$, $n_B^{eq}$ is the equilibrium value for the baryon density, and the Chern-Simons diffusion rate is 
\begin{equation}
    \Gamma_{\text{diff}} \approx 18 \alpha_W^5 T^4
\end{equation}
before the electroweak transition \cite{Burnier:2005hp,DOnofrio:2014rug}. We can estimate the timescale over which sphaleron processes restore equilibrium as 
\begin{equation}
    \tau_{\text{sph}} \approx \left(\frac{1}{n_B-n_B^{eq}} \frac{d n_B}{dt}\right)^{-1} \approx 10^5 T^{-1}
\end{equation}
It is clear that over the cooling timescale, there is more than enough time for sphaleron processes to keep particle densities in equilibrium. 

Finally, we need to consider the time taken for lepton number to diffuse out of the nugget region. Before the electroweak transition, interactions between the matter particles are all unsuppressed due to the vanishing of the W and Z boson masses. As a result, the mean free path of the matter particles is small relative to the size of the nugget and it is expected that this will slow the leaking of $B-L$ number \footnote{This is particularly important for neutrinos which, due to their long mean free path after electroweak symmetry breaking, significantly contribute to the dynamics of typical quark nuggets.}. A very rough approximation of the diffusion timescale can be made by considering the diffusion equation 
\begin{equation}
    \frac{\partial n_L}{\partial t} = D \nabla^2 n_L.
\end{equation}
In Fourier space with an infinite domain, this is solved by 
\begin{equation}
    n_L(\Vec{k},t) = n_L(\Vec{k},0) e^{- D|\Vec{k}|^2 t}.
\end{equation}
The largest physical size modes (those with the smallest $\Vec{k}$) are the modes which last the longest before diffusing away, and a shorter mean free path, corresponding to a smaller diffusion constant $D$, results in longer-lived modes. Starting with a nugget of size $R$, surrounded by no net lepton number, the longest modes will have a wavelength on the order of $4R$, with $k\approx \frac{2\pi}{4R}$. The timescale over which these modes would diffuse away would then be
\begin{equation}
    \tau_{\text{diff}} \approx \frac{2R^2}{\pi^2 D}.
\end{equation}
The radius of the quark nugget is on the order of $10^{-2} Q_\text{B-L}^{1/3}~\mathrm{MeV}^{-1}$ and the diffusion constant can be assumed to be on the order of $B^{-1/4}$. This gives a diffusion timescale on the order of $\tau_{\text{diff}}\approx 10^{-3}Q_\text{B-L}^{2/3} ~\mathrm{MeV}^{-1}$. For quark nuggets with $Q_\text{B-L}>10^{31}$, as is the case for the scenario outlined above, their diffusion timescale is longer than the cooling timescale to the chiral-electroweak phase transitions. Hence, the average quark nugget would not have evaporated significantly before this time.

\subsection{Mass, radius, and abundance of quark-lepton nuggets}
The mass and radius of the nuggets at zero temperature and with $B-L$ number, $Q_\text{B-L}$, are derived by considering the zero temperature limit of the constrained free energy in Appendix \ref{free_energy}. Assuming that the nuggets do not evaporate significantly after the chiral-electroweak transition, the mass and radius are approximately
\begin{eqnarray}
    m_\text{QN}(t_0) = \mu_\text{B-L}(0) Q_\text{B-L}(t_0) \sim  10^{9}\,\text{kg}\\
    R_\text{QN}(t_0) = \left(\frac{81\pi}{170}Q_\text{B-L}\right)^{1/3}\frac{1}{\mu_\text{B-L}(0)} \sim 1 \text{mm}
\end{eqnarray}

The quark nugget abundance today can be roughly estimated using the baryon number density today, since both quantities were dependent on the baryon density at the time of the confining transition. Following \cite{Bai:2018vik}, we begin with the ratio of the density parameters for quark nuggets and the rest of baryonic matter today, which can be expressed as a ratio of their energy densities and then their number densities
\begin{equation}
    \frac{\Omega_{QN}(t_0)}{\Omega_B(t_0)} = \frac{\rho_{QN}(t_0)}{\rho_B(t_0)} = \frac{m_{QN}(t_0) n_{QN}(t_0)}{m_p n_B(t_0)}.
\end{equation}
Since both the quark nuggets and other baryons dilute in the expanding universe as matter, their number density ratio will remain the same during expansion,
\begin{equation}
    \frac{n_{QN}(t_0)}{n_B(t_0)}=\frac{n_{QN}(t_\text{QCD})}{n_B^{(h)}(t_\text{QCD})}.
\end{equation}
Next, we can express the baryon density at the hadronisation transition in terms of the baryon density inside the nuggets with the ratio $r$ estimated in Appendix \ref{densityratio}, $n_{B}^{(h)}= r \cdot n_{B}^{(q)} = r \frac{Q_B(t_\text{QCD})}{V_{QN}(t_\text{QCD})}$, where the baryon number density in each quark phase bubble at this time can be expressed as their average total baryon number per nugget, $Q_B$ divided by the volume of the quark phase at that time, $V_{QN}(t_\text{QCD})$. Putting it all together, along with the mass of the quark nugget today, we get
\begin{equation}
    \frac{\Omega_{QN}(t_0)}{\Omega_B(t_0)} = \frac{1}{r} \frac{\mu_\text{B-L}(0)}{m_p}\frac{Q_\text{B-L}(t_0)}{Q_B(t_\text{QCD})} (n_{QN}(t_\text{QCD}) V_{QN}(t_\text{QCD})).
\end{equation}
The nugget number density times the average volume per quark phase pocket is roughly the volume fraction of the universe at the time when equilibrium is lost, which has thus far been assumed to be the percolation time. If we take this to be $n_{QN} V_{QN}\sim0.7$, assume no evaporation such that  $Q_\text{B-L}(t_0)\sim Q_B(t_\text{QCD})$ and input the previously calculated values for $\mu_\text{B-L}$ and $r$ along with the proton mass today, we get a ratio of $\frac{\Omega_{QN}(t_0)}{\Omega_B(t_0)}\sim 0.03$ which, even with the heavy approximations used, suggests that the quark-lepton nuggets in this model do not make up the majority of observed dark matter.

\section{Summary and discussions}
\label{dis}
In this paper, we have investigated the cosmological implications of the minimal Standard Model with non-linearly realised scale invariance. The model addresses the stability of the electroweak scale under radiative corrections in a technically natural manner and predicts a single additional particle in the low-energy regime: the dilaton, a pseudo-Goldstone boson associated with the non-linearly realised scale invariance. To ensure a stable hierarchy between the Planck mass and the electroweak scale, the dilaton is required to be a very light state ($m_{\chi}\sim 10^{-8}$ eV) and is feebly coupled to Standard Model particles. Consequently, the theory exhibits no detectable deviations from the Standard Model at high-energy colliders.

Remarkably, however, the early-universe evolution in this framework differs dramatically from the standard cosmological history. In particular, owing to the underlying scale invariance of the Higgs–dilaton potential, the Higgs field remains trapped in the electroweak-symmetric vacuum until the QCD chiral phase transition, which occurs at temperatures of order $\sim 28$ MeV and subsequently triggers electroweak symmetry breaking. A range of physical phenomena are expected to occur during this unorthodox phase transition, which we have studied using some simplified approximations. 

We have investigated primordial black hole (PBH) production during the QCD phase transition within our scenario, which, unlike the standard cosmological evolution, is first-order and proceeds via the nucleation of vacuum bubbles. We considered two primary mechanisms for black hole formation. The first involves the collapse of local overdensities generated by the dynamics of expanding bubbles; however, we find this mechanism to be inefficient within our framework.

Alternatively, pre-existing overdensities, presumably generated during the inflationary epoch, may collapse to form PBHs either during hadronisation or during the subsequent chiral electroweak phase transition. In both cases, a substantial PBH abundance can be produced, with characteristic masses of order $\sim 40M_{\odot}$ (hadronisation transition) and $\sim 3M_{\odot}$ (chiral electroweak transition). The resulting PBH abundance, however, is highly sensitive to the details of the underlying inflationary model.

We have also examined the production of gravitational waves during the QCD chiral phase transition and find that their amplitudes lie below the sensitivity of current and planned gravitational-wave observatories.

Finally, we studied the formation of multi-quark–lepton nuggets during the hadronisation phase transition. Our estimates indicate that a typical metastable nugget has a mass of order $\sim 10^{9}\mathrm{kg}$ and a size of order $\sim 1\mathrm{mm}$. However, their abundance constitutes only about 3\% of the total baryonic matter in the universe.

We emphasise that our results should be regarded as order-of-magnitude estimates, as they rely on a number of simplifying assumptions, which are explicitly stated in the main text. These include approximations in modelling QCD-triggered electroweak phase transitions, as well as extrapolations from previously reported numerical simulations. A more comprehensive numerical analysis of these phase transitions, using advanced computational tools \cite{Athron:2020sbe}, would therefore be highly desirable.

Nevertheless, an important and robust feature of our scenario is that the approximate scale invariance implies only a small difference in energy density between the symmetric and broken phases relative to the radiation energy density at that time. Consequently, the available latent heat that drives bubble dynamics is limited. This, in turn, leads to relatively weak signals, such as gravitational waves with small amplitudes.

While our analysis has focused on the minimal Standard Model framework with non-linearly realised scale invariance, it is plausible that short-distance physics involves extensions of the Standard Model. Such extensions could, for example, account for neutrino masses, stabilise the electroweak vacuum (see, e.g., \cite{Kobakhidze:2013pya}), or resolve the strong CP problem. In addition, as argued in \cite{Dvali:2024dlb}, new physics is to be realised within a supersymmetric framework. Extending the present analysis to such theories would be an interesting direction for future work.

\section*{Acknowledgements}
Part of this work was completed while the authors were visiting the Max Planck Institute for Physics and the Arnold Sommerfeld Centre for Theoretical Physics at Ludwig Maximilian University of Munich. We thank Gia Dvali and the members of his group for their generous hospitality and support. This work was partially funded by the Australian Research Council through the Discovery Projects grant DP220101721. JC also acknowledges support from the University of Sydney Grants-in-Aid and PRSS travel grants.

\appendix
\numberwithin{equation}{section}

\section{Parameter estimation in the linear \texorpdfstring{$\sigma$}{sigma}-model}
\label{linsigparameters}
We begin with the linear $\sigma$-model Lagrangian, including the coupling to the dilaton field
\begin{equation}\mathcal{L} = \tr(\partial_\mu\Phi^\dagger \partial^\mu\Phi) -\lambda_\sigma (\tr\Phi^\dagger\Phi - \xi_\sigma\chi^2)^2 +\lambda_\kappa \tr (\Phi^\dagger\Phi \Phi M) +h.c. \end{equation}
with 
\begin{equation}
    \Phi = \frac{\sigma}{\sqrt{2n_f}}
    \exp\!\left(\frac{\sqrt{2n_f}i\pi^a T^a}{v_\sigma}\right),
\end{equation}
which is chosen such that the kinetic terms for the $\sigma$ and $\pi^a$ fields are canonically normalised when there are $n_f$ massless quarks. \\

The first term in the potential is intended to be a phenomenological symmetry breaking potential for the modulus field $\sigma$ once scale invariance is broken and the dilaton has reached its vacuum expectation value, so that we have $\frac{v_\sigma}{2} = \xi_\sigma^{1/2}v_\chi$ and the potential becomes $\frac{\lambda_\sigma}{4}(\sigma^2-v_\sigma^2)^2$. At low energies, this term dominates the effective potential for $\sigma$ since the top quark decouples and the mass term becomes proportional to $y_s \sim 10^{-6}$ instead of $y_t \sim 1$. At high energies, the coupling to the dilaton is negligible as explained in the main text, leaving just the $\frac{\lambda_\sigma}{4}\sigma^4$ part and $\sigma$-Higgs coupling term in our effective potential. We require that this Lagrangian reduces to the non-linear $\sigma$-model as the $\sigma$ field goes to $v_\sigma$,
\begin{equation}
    \mathcal{L}  = \frac{f_\pi^2}{4}\text{Tr}\left(\partial_\mu U \partial^\mu U\right) + \kappa \text{Tr} (U M) +h.c.
\end{equation}
where $U=\exp\left(\frac{2i\pi^a T^a}{f_\pi}\right)$. By comparing the kinetic terms we arrive at $v_\sigma = \sqrt{\frac{n_f}{2}}f_\pi$ where we take the pion decay constant to be $f_\pi = 94~\mathrm{MeV}$. In the non-linear $\sigma$-model, the light pion masses are given by $m_\pi^2 = \frac{2 \kappa (m_u+m_d)}{f_\pi^2} \approx (135~\mathrm{MeV})^2$ which lets us determine 
\begin{equation}
    \lambda_\kappa = \frac{4 m_{\pi}^2}{f_\pi (m_u+m_d)}\approx 110.
\end{equation}
by matching $\kappa = \lambda_\kappa \frac{v_\sigma^3}{(2n_f)^{3/2}}$. Finally, $\lambda_\sigma$ is obtained by matching the $\sigma$-meson (f0(500) resonance) mass to our symmetry breaking potential $m_\sigma^2 \equiv \frac{d^2}{d\sigma^2} \left(\frac{\lambda_\sigma}{4}(\sigma^2-v_\sigma^2)\right)_{\sigma= v_\sigma} $ which gives 
\begin{equation}
    \lambda_\sigma = \frac{m_\sigma^2}{n_f f_\pi^2} \approx 9.4
\end{equation}
where we use $n_f=3$ to match with the experimental value for $m_\sigma$. At first glance, the large values for $\lambda_\kappa$ and $\lambda_\sigma$ seem to comprise the perturbative expansion. However, when we evaluate the tree-level potential 
\begin{eqnarray}
V_0(h,\sigma) &=& 
\frac{\lambda_\sigma}{4}\sigma^4 - \frac{\lambda_\kappa y_t}{2 n_f^{3/2}}\sigma^3 h \approx 2.4 \sigma^4 - 3.7 \sigma^3 h  \nonumber \\
&=&\left(\frac{\lambda_\sigma}{4} \sin^4\theta - \frac{\lambda_\kappa y_t}{2 n_f^{3/2}}\sin^3\theta\cos\theta\right)\rho^4 \approx -0.3 \rho^4,
\end{eqnarray}
along the minimised direction of the potential, the effective coupling remains small.

\section{Chiral-electroweak transition wall velocity}
\label{bubblespeed}
The wall velocity of the expanding bubbles in a first order phase transition gives an indication of where the energy of the transition is distributed and so the dominant source of gravitational waves. The investigation of the wall velocity here closely follows the work done in \cite{Megevand:2013hwa}. We begin with the equations of motion for the modulus $\rho$-field,
\begin{equation}
    \partial_\mu \partial^\mu\rho +\frac{\partial}{\partial\rho} V_{\text{eff}}(\rho,0) +\sum_i g_i \frac{dm_i^2}{d\rho}\int \frac{d^3p}{(2\pi)^32E_i}f_i(p,z)=0
\end{equation}
where the friction term is derived using the WKB approximation for the wavefunctions of particles in the plasma, assuming that the background $\rho$ field varies slowly relative to their typical momentum. $z$ is the distance from the centre of the wall in the wall's frame of reference (the centre is defined to be such that $\int_{-\infty}^{\infty} dz\,z \left(\frac{d\rho}{dz}\right)^2=0 $). Multiplying by $\frac{d\rho}{dz}$ and integrating over the wall 
\begin{equation}\label{walleom}\left(\int_{z_-}^{z_+} \left(\frac{d\rho}{dz}\right)^2 dz\right)\ddot{z}_w = \int_{z_-}^{z_+} dz\frac{\partial V_{\text{eff}}(\rho,0)}{\partial\rho}\frac{d\rho}{dz} +\int_{z_-}^{z_+} dz \frac{d\rho}{dz}\sum_i g_i \frac{dm_i^2}{d\rho}\int \frac{d^3p}{(2\pi)^32E_i}f_i(p,z),
\end{equation}
gives an equation for the acceleration of the centre of the wall $\ddot{z}_w$. The first term on the right hand side is a driving force per unit area, $\frac{F_{\text{dr}}}{A}$, due to the differences in vacuum pressure between the two phases while the second term can be seen as a friction term, $\frac{F_{\text{fr}}}{A}$, resulting from the interactions of the wall with the particles in the plasma, specifically the change in mass of interacting particles as they cross the wall. Summing over each species of particle in the plasma and integrating over their, as of yet unknown, thermal distributions gives the net pressure on the wall. There are two limits for the bubble wall velocity in which the thermal distributions of particles can be easily estimated: an ultra-relativistic, runaway limit in which the bubble wall continuously accelerates and a non-relativistic limit in which the wall reaches some non-relativistic terminal velocity, both of which need to be checked.
\subsection{Runaway limit}
In the runaway limit, the incoming particles have no time to equilibrate, and so their distributions remain in what was thermal equilibrium (simplified to be a Bose-Einstein distribution) in front of the wall, $f_i^{\text{eq.}}(p,z_+) = \frac{1}{e^{\sqrt{p^2+m_{i+}^2}/T_p^{(\chi)}}-1}$, (in the plasma's frame of reference) with $m_{i+}$ their mass before the phase transition and $T_p^{(\chi)}$ the percolation temperature at the time of the transition. The integral over momenta can be evaluated in the plasma's frame of reference since it is Lorentz invariant $\int \frac{d^3p}{(2\pi)^32E_i}f_i^{\text{eq+}}(p,z) \equiv \frac{T_p^{(\chi)}{}^2}{2}c_4\left(\frac{m_{i+}^2}{T_p^{(\chi)}{}^2}\right)$ with 
\begin{equation}
    c_4(x)\equiv \frac{1}{2\pi^2}\int_{\sqrt{x}}^\infty dy \sqrt{y^2-x}\frac{1}{e^y-1}.
\end{equation}
The driving force on the wall is evaluated to be  
\begin{equation}\frac{F_{\text{dr}}}{A} = \int_-^+ dz\frac{\partial V_{\text{eff}}(\rho,0)}{\partial\rho}\frac{d\rho}{dz} = V_{\text{eff}}(\rho_+,0)-V_{\text{eff}}(\rho_-,0) = 5.9 \times 10^5~\mathrm{MeV}^4\end{equation}
while the friction force is 
\begin{align}
    \frac{F_{\text{fr}}}{A} &= +\int_-^+ dz \frac{d\rho}{dz}\sum_i g_i \frac{dm_i^2}{d\rho}\int \frac{d^3p}{(2\pi)^32E_i}f_i^{\text{eq+}}(p,z)\nonumber\\
    &= -\frac{T_p^{(\chi)}{}^2}{2}\sum_i g_i (m_{i-}^2-m_{i+}^2)c_4\left(\frac{m_{i+}^2}{T_p^{(\chi)}{}^2}\right)\nonumber\\
    &= -23.4 \times 10^5~\mathrm{MeV}^4
\end{align}
Given that the friction force is larger than the driving force, the bubble wall must reach some terminal velocity before getting to this runaway limit.\\

For ease of comparison to the non-relativistic limit in the next section, we redefine what we call the driving and friction forces in the runaway regime by adding and subtracting the finite temperature contribution to the effective potential to get
\begin{equation}\frac{F_{dr}}{A} = \int_-^+ dz\frac{\partial V_{\text{eff}}(\rho,T_p^{(\chi)}) }{\partial\rho}\frac{d\rho}{dz} = 1.1 \times 10^5~\mathrm{MeV}^4\end{equation}
and 
\begin{equation}\frac{F_{fr}}{A} = V_T(\rho_-,T_p^{(\chi)})-V_T(\rho_+,T_p^{(\chi)}) -\frac{T_p^{(\chi)}{}^2}{2}\sum_i g_i (m_{i-}^2-m_{i+}^2)c_4\left(\frac{m_{i+}^2}{T_p^{(\chi)}{}^2}\right)=-18.7 \times 10^5~\mathrm{MeV}^4\end{equation}

\subsection{Non-relativistic limit}
This limit relies on a small wall velocity leading to small deviations from equilibrium $f_i(p,z) = f_i^{\text{eq.}}(p,T)+\delta f_i(p,z)$. The momentum integral over the equilibrium distribution gives the 1-loop contribution to the effective potential 
\begin{equation}
    \sum_i g_i \frac{dm_i^2}{d\rho}\int \frac{d^3p}{(2\pi)^32E_i}f_i^{\text{eq}}(p,z) = \frac{\partial V_T}{\partial\rho}
\end{equation}
Including this in the previous expression for the driving force gives
\begin{equation}\frac{F_{dr}}{A} = \int_-^+ dz\frac{\partial V_{\text{eff}}(\rho,T_p^{(\chi)}) }{\partial\rho}\frac{d\rho}{dz} = 1.1 \times 10^5~\mathrm{MeV}^4\end{equation}
as above, leaving a friction term which only depends on the deviations from equilibrium,
\begin{equation}
\label{NRfriction}
    \frac{F_{fr}}{A}=\int_-^+ dz \frac{d\rho}{dz}\sum_i g_i \frac{dm_i^2}{d\rho}\int \frac{d^3p}{(2\pi)^32E_i}\delta f_i(p,z).
\end{equation}
The size and profile of these deviations can be estimated using the Boltzmann equation and an ansatz for the full distributions. Assuming no variation in the $x$ or $y$ directions,
\begin{equation}
    \partial_t f_i +\dot{Z}\partial_Z f_i +\dot{p}_z\partial_{p_z}f_i = -C[\{f_j\}]
\end{equation}
where Z is the longitudinal coordinate in the plasma's frame of reference and $C[\{f_i\}]$ is a collision functional of the distributions of each species in the plasma. Integrating over the momentum, $\int \frac{d^3p}{(2\pi)^3}$, cancels the second and third terms on the left-hand side as they each end up being odd in $p_z$. In the wall's frame of reference, $\partial_t \rightarrow -v_w \partial_z$ ($z$ being the longitudinal coordinate in the wall's frame of reference), which results in
\begin{equation}
\label{boltzmann}
    \int \frac{d^3p}{(2\pi)^3}\partial_z f_i  = \frac{1}{v_w}\int \frac{d^3p}{(2\pi)^3}C[\{f_j\}].
\end{equation}
Given a sufficiently slow moving wall, the deviation from equilibrium will be small enough that it can be approximated as another equilibrium distribution of particles with slightly smaller energies, $f_i(\frac{E_i}{T}) = f_i^{\text{eq}}(\frac{E_i}{T}-\delta_i(z))$, with the further assumption that the temperature of the plasma does not change significantly across the width of the wall, that is, that the majority of the latent heat goes into bulk motion of the fluid. Inserting the ansatz $f_i^{\text{eq}}(\frac{E_i}{T}-\delta_i)$, taking the partial derivative with respect to $z$, expanding to first order in $\delta_i(z)$ and then changing variables to $x_i\equiv \frac{m_i^2(\rho(z))}{T^2}$ results in the first order differential equation
\begin{equation}
\label{delta}
    2 c_3(x_i) \frac{d\delta_i}{dx_i} + \left(c_2(x_i)+\frac{2\Gamma_i}{v_w \frac{dx_i}{dz}}\right)\delta_i-c_1(x_i) = 0.
\end{equation}
with
\begin{align}
    c_1(x)&\equiv \frac{-1}{2\pi^2}\int_{\sqrt{x}}^\infty dy \sqrt{y^2-x}\frac{e^y}{(e^y-1)^2}\nonumber\\
    c_2(x)&\equiv \frac{1}{2\pi^2}\int_{\sqrt{x}}^\infty dy \sqrt{y^2-x}\frac{3e^y-e^{2y}}{(e^y-1)^3}\nonumber\\
    c_3(x)&\equiv \frac{-1}{2\pi^2}\int_{\sqrt{x}}^\infty dy y\sqrt{y^2-x}\frac{e^y}{(e^y-1)^2}
\end{align}
which come from taking momentum integrals of the equilibrium distribution function and its derivatives with respect to the mass of the particles, $x_i$. For the collision term, the zeroth order, equilibrium term should not contribute to a change in particle number so we simplify with $\int \frac{d^3p}{(2\pi)^3}C[\{f_j\}]=T^3 \Gamma_i \delta_i$ such that in the limit of a stationary wall, there will be no deviations and the system will stay in thermal equilibrium. The constants, $\Gamma_i$, are estimated based on the dominant inelastic interaction processes for each particle species. If we take the collision integral to be made up of a single two-particle self-annihilation decay process (ignoring Bose enhancement and Pauli blocking), we would have
\begin{equation}
    -T^3 \delta_1 \Gamma_1 = \frac{1}{2}\int \prod_{i=1}^4 \frac{d^3p_i}{(2\pi)^3} (2\pi)^4\delta^{(4)}(p_1+p_2+p_3+p_4)|\mathcal{M}_{12\rightarrow34}|^2 (f_3f_4-f_1f_2)
\end{equation}
Assuming this is the dominant process, the equilibrium parts of $f_1$, $f_2$, $f_3$ and $f_4$ should cancel one another, leaving $f_3 f_4-f_1 f_2 \approx -\delta_3 \frac{d f_3^{\text{eq}}}{d E_3}f_4^{\text{eq}}-\delta_4 \frac{d f_4^{\text{eq}}}{d E_4}f_3^{\text{eq}}+\delta_2 \frac{d f_2^{\text{eq}}}{d E_2}f_1^{\text{eq}}+\delta_1 \frac{d f_1^{\text{eq}}}{d E_1}f_1^{\text{eq}}+\mathcal{O}(\delta^2)$. The dominant contribution to the friction term will come from particles strongly coupled to the Higgs and sigma fields, which gain a large mass as they cross the bubble wall. These particles experience a larger force from the wall and can be expected to be further out of equilibrium, hence for a self-annihilation decay process, we assume $\delta_1=\delta_2\gg \delta_3,\, \delta_4$, which gives 
\begin{equation}
    -T^3 \delta_1 \Gamma_1 = \delta_1 \int \prod_{i=1}^4 \frac{d^3p_i}{(2\pi)^3} (2\pi)^4\delta^{(4)}(p_1+p_2+p_3+p_4)|\mathcal{M}_{12\rightarrow34}|^2 f_1^{\text{eq}}\frac{d f_1^{\text{eq}}}{d E_1}.
\end{equation}
Very roughly estimating the momentum integral over the thermal distributions and assuming the amplitude has no momentum dependence gives
\begin{equation}
    \Gamma_i \approx |\mathcal{M}_{ii\rightarrow jk}|^2 T
\end{equation}
The heavy pions and W and Z bosons contribute the most to the friction term since they have large changes in mass and many degrees of freedom. The dominant pion decay is to a $\sigma$ meson and Higgs boson with tree-level amplitudes $\mathcal{M}_{\pi_{35}\pi_{35}\rightarrow\sigma h}=\frac{5}{3}\frac{\lambda_\kappa y_t}{4\sqrt{n_f}}$ and $\mathcal{M}_{\pi_{25-34}\pi_{25-34}\rightarrow\sigma h}=\frac{\lambda_\kappa y_t}{4\sqrt{n_f}}$, while the Ws and Zs predominantly decay with amplitudes $\mathcal{M}_{W\rightarrow \psi \psi}=\frac{g}{2}$ and $\mathcal{M}_{Z\rightarrow \psi \psi}=\frac{\sqrt{g^2+g'{}^2}}{2}$, respectively.\\

Solving Eq. (\ref{delta}) also requires the profile of the bubble wall in $\frac{dx_i}{d z}$. This was estimated by fitting an ansatz to the equations of motion for the $\rho$ field that smoothly interpolates between the vacuum in front of the wall, $\rho_+$, and behind it, $\rho_-$.
\begin{equation}
    \rho(z,L_w) = \frac{\rho_-+\rho_+}{2} +\frac{\rho_--\rho_+}{2}\tanh \left(\frac{z}{L_w}\right)
\end{equation}
By minimising the error in the equations of motion for $\rho(z, L_w)$, $\partial_\mu \partial^\mu\rho +\frac{\partial}{\partial\rho} V_{\text{eff}}(\rho, T)=0$ (where I've neglected the friction term which is negligible in the non-relativistic case) we get a wall width of $L_w=0.03 ~\mathrm{MeV}^{-1}$ which can then be used to estimate $\frac{dx}{dz}(x)$.\\

Equation (\ref{delta}) is solved numerically for each of the $\delta_i$ corresponding to the most strongly coupled particles, with the boundary condition $\delta_i(x_+) = 0$ so that there is no deviation from equilibrium far in front of the wall. Then, the friction term becomes 
\begin{align}
    \frac{F_{fr}}{A}&=-\int_-^+ dz \frac{d\rho}{dz}\sum_i g_i \frac{dm_i^2}{d\rho}\int \frac{d^3p}{(2\pi)^32E_i}\delta_i(x_i) \frac{df_i^{\text{eq}}(E_i,x_i)}{d E_i}\nonumber \\
    &= \frac{-T^4}{2}\sum_i g_i \int_-^+ dx\, \delta_i(x_i) c_{1}(x_i)
\end{align}
which was also evaluated numerically. For small wall velocities, we find that $\frac{F_{fr}}{A}\approx -1800 v_w ~\mathrm{MeV}^4$.

Clearly, $\frac{F_{fr}}{A}$ is always much less than $\frac{F_{dr}}{A}$ even up to $v_w=1$ (which long past its domain of applicability). Given that there is no speed for which the non-relativistic limit produces balanced forces on the wall, the wall must reach speeds at which the approximations used break down but is not able to reach the runaway limit. 

\subsection{Interpolating between the limits}
Following the method proposed in \cite{Megevand:2013hwa}, we adopt an interpolating friction term in the equations of motion of the form 
\begin{equation}
    \partial_\mu\partial^\mu\rho +\frac{\partial V_{\text{eff}}}{\partial\rho} -\frac{\partial_\mu \rho\, u^\mu \,f(\rho)}{\sqrt{1+(\partial_\mu \rho\, u^\mu \,g(\rho))^2}}=0,
\end{equation}
where $f(\rho)$ and $g(\rho)$ are arbitrary functions chosen to match the friction force in both of the limits $v_w\rightarrow1$, $\gamma(v_w)\rightarrow \infty$ and $v_w\rightarrow0$, $\gamma(v_w)\rightarrow 1$ and $u^\mu$ is the fluid 4-velocity of the surrounding plasma. Assuming a constant fluid velocity, this equation of motion leads to the friction force on the bubble
\begin{equation}
    \frac{F_{fr}}{A}(v_w) = -\int_-^+ d\rho\, \frac{\partial_z\rho \,v_w \gamma(v_w) f(\rho) }{\sqrt{1+(\partial_z\rho \,v_w \gamma(v_w) g(\rho))^2}}.
\end{equation}
We redefine $f(\rho)$ and $g(\rho)$ to absorb the $\partial_z\rho$ dependence, then set
\begin{equation}
    f(\rho) = \frac{T^2}{2}\sum_i g_i \frac{dm_i^2}{d\rho}\delta_i(x)c_{1,i}(x)
\end{equation}
and 
\begin{equation}
    \frac{f(\rho)}{g(\rho)} = \frac{T^2}{2}\sum_i g_i \frac{dm_i^2}{d\rho}\left(\frac{-1}{\pi^2}\frac{d J_B(x)}{dx}+c_4(x_+)\right)
\end{equation}
to fit the two limits. Using this interpolating friction term, the terminal velocity for which $\frac{F_{fr}}{A}(v_w)=-\frac{F_{dr}}{A}$ was found to be $v_w=0.99987$ with a gamma factor $\gamma(v_w)\approx 60$.

\section{Baryon number density ratio at the confining transition}
\label{densityratio}
To calculate the baryon number densities in the two phases, we assume thermal and chemical equilibrium between the two phases and treat the fermions in each phase as a Fermi gas. The fermion densities are given by 
\begin{equation}
    n_f = g_f \int \frac{d^3p}{(2\pi)^3}\, \frac{1}{e^{\epsilon-\mu_f/T}+1}
\end{equation}
where $g_f$ are the internal degrees of freedom for that particle type, and the chemical potentials for each species are set by the chemical potentials of the conserved charges, $\mu_f = \sum_i q_f^i \mu_i$. Given that sphaleron processes are active at this time, baryon and lepton numbers are not conserved separately; we instead need to consider $\mu_\text{B-L}$. We might also expect to consider hypercharge as a relevant conserved charge however, due to the delayed electroweak transition and the scale-invariant Higgs potential, the Higgs is effectively massless at this time, and so the minimum energy cost to add a Higgs particle into the system is far smaller than the energy cost to add a fermion. As a result, $\mu_Y \ll \mu_\text{B-L}$, so we can neglect the hypercharge chemical potential when computing fermion densities. In chemical equilibrium, particles related by CP conjugation have opposite chemical potentials, $\mu_{\bar{f}} = - \mu_f$, which allows us to write the net particle number density as 
\begin{equation}
    n_{f-\bar{f}}\equiv n_{f}-n_{\bar{f}} = g_f \int \frac{d^3p}{(2\pi)^3}\, \frac{\sinh{\frac{\mu_f}{T}}}{\cosh{\frac{\epsilon}{T}}+\cosh{\frac{\mu_f}{T}}}.
\end{equation}
Before the $B-L$ charge is concentrated into a quark nugget, the chemical potential is far smaller than the temperature, so we can expand in $\frac{\mu_f}{T}$,
\begin{equation}
    n_{f-\bar{f}} \approx g_f \frac{\mu_f}{T} \int \frac{d^3p}{(2\pi)^3}\, \frac{1}{\cosh{\frac{\epsilon}{T}}+1} = g_f \mu_f T^2 f(\frac{m}{T})
\end{equation}
where $f(y) = \int_0^{\infty}\frac{x^2 dx}{2\pi^2}\frac{1}{\cosh{\sqrt{x^2+y^2}}+1}$,  $f(0)=1/6$ and we've assumed a free Fermi gas so that the energy of a state with momentum $p$ is $\epsilon=\sqrt{p^2+m^2}$.

The baryon number density is then a sum over each fermion number density multiplied by their baryon number charge. For six massless quark flavours in the quark phase we get $n_B^{(q)} = \sum_f q_f^B n_{f-\bar{f}}^{(q)} = \frac{2}{3}\mu_\text{B-L}T^2$, and for $N_B$ degenerate lowest mass baryons in the hadron phase, $n_B^{(h)} = \sum_f q_f^B n_{f-\bar{f}}^{(h)} = 2 N_B f(\frac{m_p}{T})\mu_\text{B-L}T^2$. For six flavours of quark with spin $1/2$, we look at the symmetric part of the $(\mathbf{6}\otimes \mathbf{2})^3$ representation of $\text{SU}(6)\otimes \text{SU}(2)$, which has $N_B = 70$. We also need the baryon mass in this theory which we estimate by rescaling the baryon mass of the three-flavour theory, $m_p^{(6)} = m_p^{(3)}\frac{\Lambda_{\text{QCD}}^{(6)}}{\Lambda_{\text{QCD}}^{(3)}} = 938~\mathrm{MeV}\frac{89~\mathrm{MeV}}{332~\mathrm{MeV}}= 251~\mathrm{MeV}$. This gives a ratio 
\begin{equation}
    r = \frac{n_B^{(h)}}{n_B^{(q)}} = 3N_Bf(\frac{m_p}{T_c^{(h)}}) \approx 5.1 \,.
\end{equation}
Unlike in \cite{Bai:2018vik}, which had $r = 0.22$, here the baryon number density is larger in the hadron phase, so baryon number isn't being concentrated during this period of equilibrium. The difference is largely due to our smaller estimate for $m_p/T_c^{(h)}$, due to the rescaling of $m_p$.

\section{Free energy of the quark-lepton nugget}
\label{free_energy}
To model the quark-lepton nugget, we consider a volume $V$ containing some spherical region of radius $R$, within which is a quark-phase plasma at temperature $T$ and vacuum energy density $B$ and surrounded by a hadron-phase plasma at the same temperature. Within the nugget, there is some net conserved $B-L$ number, which, for now, we assume cannot leak out of the nugget. With fixed temperature and volume, the relevant thermodynamic potential to minimise is the free energy,
\begin{equation}
    F = \left(V-\frac{4\pi R^3}{3}\right) \mathcal{F}_h + \frac{4\pi R^3}{3} \mathcal{F}_q,
\end{equation}
with $\mathcal{F}_q$ and $\mathcal{F}_h$ the free energy densities in the quark and hadron phases, respectively. Conserving $B-L$ number is achieved by introducing a Lagrange multiplier $\mu_\text{B-L}$, giving 
\begin{equation}
    \Omega = \left(V-\frac{4\pi R^3}{3}\right) \mathcal{F}_h + \frac{4\pi R^3}{3} \mathcal{F}_q - \mu_\text{B-L}\left(\frac{4\pi R^3}{3} n_\text{B-L}-Q_\text{B-L}\right)
\end{equation}
as the thermodynamic potential to be minimised. The hadron phase free energy density is given by the negative of the pressure of the relativistic gas, $\mathcal{F}_h =  \frac{-\pi^2}{90}g_*^{(h)}T^4$. The free energy density inside the nugget will consist of a contribution from the relativistic bosons $ \frac{-\pi^2}{90}g_*^{(q\, \text{bosons})}T^4$, the vacuum energy density $B$ and the fermions, modelled here as a non-interacting Fermi gas, which can be calculated from their energy and entropy densities, $\mathcal{F} = \mathcal{E}-Ts$, 
\begin{align}
    \mathcal{E}(\mu_f,T) &= \sum_f g_f \int \frac{d^3p}{(2\pi)^3} \epsilon \frac{1}{e^{-(\epsilon-\mu_f)/T}+1}\\
    Ts(\mu_f,T) &= \sum_f g_f \int \frac{d^3p}{(2\pi)^3} (\epsilon-\mu_f)\frac{1}{e^{-(\epsilon-\mu_f)/T}+1} + T \log\left(1+e^{-(\epsilon-\mu_f)/T}\right).
\end{align}
The total $B-L$ number density is  
\begin{equation}
    n_\text{B-L}(\mu_f,T) = \sum_f g_f \int \frac{d^3p}{(2\pi)^3} q_\text{B-L}^f \frac{1}{e^{-(\epsilon-\mu_f)/T}+1}
\end{equation}
where $q_\text{B-L}^f$ is the $B-L$ charge of fermion species $f$ and $\mu_f$ is its chemical potential. Putting it all together, the constrained free energy is 
\begin{align}
    \Omega  &= -(V-\frac{4\pi R^3}{3}) \frac{\pi^2}{90}g_*^{(h)}T^4 +\mu_\text{B-L} Q_\text{B-L} + \frac{4\pi R^3}{3}\bigg(- \frac{\pi^2}{90}g_*^{(q\, \text{bosons})}T^4 + B \nonumber\\
    &+ \sum_f g_f \int \frac{d^3p}{(2\pi)^3}\big( -T \log\left(1+e^{-(\epsilon-\mu_f)/T}\right)-T \log\left(1+e^{-(\epsilon+\mu_f)/T}\right) \nonumber\\
    &+ (\mu_f-\mu_\text{B-L} q_\text{B-L}^f) \frac{\sinh(\mu_f/T)}{\cosh(\epsilon/T)+\cosh(\mu_f/T)}\big)\bigg)
\end{align}
where the sum over species $f$ now just runs over particle species, taking into account that $\mu_{\bar{f}} = -\mu_f$ where the species $f$ and $\bar{f}$ are related via CP conjugation, allowing for the combination of Fermi-Dirac distributions into the hyperbolic trig term. This potential should be stationary with respect to each $\mu_f$, $\mu_\text{B-L}$, and R, which gives a set of coupled equations. Firstly, $\frac{\partial \Omega}{\partial \mu_\text{B-L}}=0$ gives the conservation of $B-L$ number 
\begin{equation}
    Q_\text{B-L} - \frac{4\pi R^3}{3}\sum_f g_f \int \frac{d^3p}{(2\pi)^3} q_\text{B-L}^f \frac{\sinh(\mu_f/T)}{\cosh(\epsilon/T)+\cosh(\mu_f/T)}=0
\end{equation}
which, for the massless fermions inside the nugget, can be integrated to 
\begin{equation}
\label{conservedcharge}
    Q_\text{B-L} =\frac{4\pi R^3}{3}\sum_f g_f q_\text{B-L}^f  \frac{1}{6\pi^2} \mu_f (\pi^2 T^2 + \mu_f^2).
\end{equation}
Next, for each fermion species, we have $\frac{\partial \Omega}{\partial \mu_f}=0$,
\begin{equation}
    \int\frac{d^3p}{(2\pi)^3} (\mu_f-\mu_\text{B-L} q_\text{B-L}^f) \frac{\partial}{\partial \mu_f}\left[\frac{\sinh(\mu_f/T)}{\cosh(\epsilon/T)+\cosh(\mu_f/T)}\right] = 0
\end{equation}
which for arbitrary temperatures, implies $\mu_f = \mu_\text{B-L} q_\text{B-L}^f$. Substituting this back into the expression for $\Omega$ and performing the integral over the log terms gives 
\begin{align}
    \Omega  &= - V \frac{\pi^2}{90}g_*^{(h)}T^4 +\mu_\text{B-L}(R,Q_\text{B-L},T) Q_\text{B-L} \\
    &+ \frac{4\pi R^3}{3}\left(\frac{\pi^2}{90}g_*^{(h)}T^4-\frac{\pi^2}{90}g_*^{(q)}T^4 + B - \frac{1}{24\pi^2}\sum_f g_f \left(\mu_f(R,Q_\text{B-L},T)^4 + 2\pi^2 T^2 \mu_f(R,Q_\text{B-L},T)^2\right)\right) \nonumber
\end{align}
where $g_*^{(q)}$ now includes all the relativistic degrees of freedom, including the fermions and $\mu_\text{B-L}(R, Q_\text{B-L}, T)$ is fixed by Eq. (\ref{conservedcharge}). Finally, $\frac{d \Omega}{d R}=0$,
\begin{align}
    &\frac{d \mu_\text{B-L}}{dR} Q_\text{B-L} + \frac{4\pi R^3}{3}\left(-\frac{1}{6\pi^2}\sum_f g_f (\mu_f^2 + \pi^2 T^2)\mu_f \frac{d \mu_f}{dR}\right) \nonumber \\
    &+4\pi R^2\left(\frac{\pi^2}{90}g_*^{(h)}T^4-\frac{\pi^2}{90}g_*^{(q)}T^4 + B - \frac{1}{24\pi^2}\sum_f g_f (\mu_f^4 + 2\pi^2 T^2 \mu_f^2)\right) = 0
\end{align}
Substituting the expression for $Q_\text{B-L}$ in terms of $R$ and $\mu_f$ using Eq. (\ref{conservedcharge}) and also using $\frac{d \mu_f}{dR} =  q_\text{B-L}^f\frac{d\mu_\text{B-L}}{dR}$, the first two terms cancel, leaving just the second line. This happens to just be the condition that the pressures on either side of the nugget wall should balance and is simply a quartic equation in $\mu_\text{B-L}$. Solving it gives 
\begin{equation}
\label{chempot}
    \mu_\text{B-L}(T) = \sqrt{-\beta +\sqrt{\beta^2+\gamma}}
\end{equation}
with $\beta = \frac{117\pi^2}{85}T^2$ and $\gamma = \frac{216\pi^2}{85}(B-\frac{\pi^2}{90}(g_*^{(q)}-g_*^{(h)})T^4)$. We can also solve Eq. (\ref{conservedcharge}) to find the radius of the nugget
\begin{equation}
    R(Q_\text{B-L},T) = \left(\frac{9\pi}{2}\frac{Q_\text{B-L}}{13\pi^2 T^2 \mu_\text{B-L}(T)^2+\frac{85}{9}\mu_\text{B-L}(T)^3}\right)^{1/3}.
\end{equation}
Substituting the pressure balance relation, we arrive at a final expression for the constrained free-energy,
\begin{equation}
    \Omega  = - P(T) V  +\mu_\text{B-L}(T) Q_\text{B-L} \\
\end{equation}
where $P(T) = \frac{\pi^2}{90}g_*^{(h)}T^4$ is the pressure throughout the system, balanced on either side of the phase transition walls, and the temperature-dependent chemical potential is given by Eq. \ref{chempot}.

\end{document}